\documentclass[secnumarabic, graphics,floatfix, nofootinbib,tightenlines,nobibnotes, aps, prl, 12pt]{revtex4-1}
\usepackage{graphicx}
\usepackage{amsmath,amssymb}
\usepackage{amsfonts}
\usepackage{multirow}

\begin{document}
\title{A mesoscopic approach on stability and phase transition between different traffic flow states}

\author{Wei-Liang Qian$^{1,2}$, Bin Wang$^{3}$, Kai Lin$^{4}$, Romuel F. Machado$^{5}$ and Yogiro Hama$^{6}$}
\affiliation{$^{1}$Escola de Engenharia de Lorena, Universidade de S\~ao Paulo, SP, Brasil}
\affiliation{$^{2}$Faculdade de Engenharia de Guaratinguet\'a, Universidade Estadual Paulista, SP, Brasil}
\affiliation{$^{3}$Shanghai Jiaotong University, Shanghai, China}
\affiliation{$^{4}$Instituto de F\'isica e Qu\'imica, Universidade Federal de Itajub\'a, MG, Brasil}
\affiliation{$^{5}$Instituto de Ci\^encias Exatas e Biologicas, Universidade Federal de Ouro Preto, SP, Brasil}
\affiliation{$^{6}$Instituto de F\'isica, Universidade de S\~ao Paulo, SP, Brasil}
\date{\today}

\begin{abstract}
It is understood that congestion in traffic can be interpreted in terms of the instability of the equation of dynamic motion. The evolution of a traffic system from an unstable or metastable state to a globally stable state bears a strong resemblance to the phase transition in thermodynamics. In this work, we explore the underlying physics of the traffic system, by examining closely the physical properties and mathematical constraints of the phase transitions therein. By using a mesoscopic approach, one entitles the catastrophe model the same physical content as in the Landau's theory, and uncovers its close connections to the instability of the equation of motion and to the transition between different traffic states. In addition to the one-dimensional configuration space, we generalize our discussions to the higher-dimensional case, where the observed temporal oscillation in traffic flow data is attributed to the curl of a vector field. We exhibit that our model can reproduce the main features of the observed fundamental diagram including the inverse-$\lambda$ shape and the wide scattering of congested traffic data. When properly parameterized, the main feature of the data can be reproduced reasonably well either in terms of the oscillating congested traffic or in terms of the synchronized flow.
\end{abstract}

\maketitle

\section{I. Introduction}

Traffic flow modeling has long attracted the attention of physicists (for reviews, see for examples \cite{traffic-flow-review-07,traffic-flow-review-01,traffic-flow-btz-02,traffic-flow-review-02,traffic-flow-review-03,traffic-flow-review-04}). In particular, it is understood by many authors that the traffic congestion is closely connected to the instability of the equation of motion (EoM) of the corresponding traffic theory \cite{traffic-flow-hydrodynamics-04,traffic-flow-hydrodynamics-12,traffic-flow-micro-07,traffic-flow-micro-08,traffic-flow-micro-09}. For instance, by using a microscopic car-following model, Bando et al. \cite{traffic-flow-micro-07} investigated the imaginary part of the frequency of small oscillations around a given steady state. It turns out that the signature of the imaginary frequency indicates the properties of linear stability of the steady state solution: negative imaginary frequency implies that small perturbation will blow up exponentially to render the system unstable. For finite size perturbations, one usually has to resort to numerical techniques. By numerical calculations, the authors showed that finite size perturbation will cause the vehicles to cluster into congestion, and further study showed \cite{traffic-flow-micro-09} that in this specific model the traffic congestion is related to the continuously deformable kink soliton solution. Another study was carried out by Kerner and Konhauser \cite{traffic-flow-hydrodynamics-04} from a hydrodynamical point of view. Cluster formation was encountered numerically where small deviations about homogeneous steady traffic flow destabilize the system. Although the above approaches are different in their respective levels of aggregation, they are very similar in nature since they all follow the general arguments used to investigate the stability in a classical dynamical system \cite{ordinary-differential-equation-teschl}. 

When the system is locally unstable for infinitesimal oscillations, it is not known beforehand whether finite deviation from a steady state may lead to global instability and consequently to any unphysical situation. For a realistic approach, the system must always be bounded and therefore, divergence shall not occur to any physical observable. For instance, quantities such as the total number of vehicles, vehicle velocity and acceleration must always remain finite; in other words, the system can be linearly unstable, but the amplitude shall never increase unboundedly. In fact, such physical constraints can be seen as necessary conditions for a realistic model. If one follows this train of thought, it is noted that for a thermodynamical system the phase transition is usually caused by perturbations at locally unstable or metastable states. Under such situation, the system evolves and eventually relocates itself to another locally or globally stable state; at the same time, the thermodynamical system in question must always be bounded. To determine if one state is more favorable than another, usually the thermodynamical potential is introduced, so that the locally stable and metastable states are related to the local minima while the globally stable state corresponds to the global minimum of the potential. The well-known Landau's theory of phase transition \cite{statistical-mechanics-landau} is one of such phenomenological approaches dealing with transitions between different phases via the concept of {\it free energy}. 

Incidentally, in the traffic flow theory the {\it catastrophe model} \cite{traffic-flow-catastrophe-01} bears an analogy with the Landau's theory. The theory was first developed by Thom \cite{traffic-flow-catastrophe-03} and Zeeman \cite{traffic-flow-catastrophe-04} and was used mostly to model catastrophic events. To describe the observed discontinuous transition between different traffic states, it was later applied to the field of traffic flow. To this end, a potential function is introduced, where the minima of the function correspond to the observed traffic states. In the case of the {\it cusp catastrophe} model \cite{traffic-flow-catastrophe-02}, the extrema of the function are single-valued on one side but three folded on the other. The folded surface in the parameter space as well as the consequent occurrence of {\it sudden jump} of observables featured in this model are indeed very similar to those in the theory of phase transition. Despite its success, the catastrophe model does not have a microscopic origin; it attempts neither to deal with the temporal evolution of the traffic system, nor to derive the form of the potential function from first principle. Naturally, there is no concept such as temperature or thermodynamical ensemble in the catastrophe approach, so the physical content of the model seems to be very different from that of Landau's theory of phase transition. Nevertheless, the main idea proposed by the catastrophe model is intriguing owing to its unique viewpoint to the problem.

Over the last decade, the investigation of traffic flow was advanced by more realistic approaches featuring parametrization based on observed traffic data, vehicle models as well as government regulations. Many ingenious works \cite{traffic-flow-hydrodynamics-04,traffic-flow-hydrodynamics-07,traffic-flow-hydrodynamics-15,traffic-flow-hydrodynamics-13,traffic-flow-hydrodynamics-14,traffic-flow-hydrodynamics-12,traffic-flow-micro-17} were carried out aiming at a deeper insight into the traffic system; in particularly, the study of the different phases of the traffic system and their coexistence and transitions. As an example, in \cite{traffic-flow-micro-17}, a microscopic intelligent driver model (IDM) was proposed and used to explore various phases of traffic flow. By numerical simulations, the model was shown to be very successful in reproducing the observed data once the boundary conditions of the simulated system were properly given. For a homogeneous system, both the free traffic (FT) state and the homogeneous congestion traffic (HCT) state were reproduced by the model and are shown to be linearly stable against small perturbations. What makes the model more significant is when it is applied to study traffic systems with inhomogeneity. The authors show that the model is capable of reproducing other non-stationary traffic states, in particular, the oscillatory states observed in the data \cite{traffic-flow-data-01,traffic-flow-data-04,traffic-flow-data-05,traffic-flow-data-06} which are characterized by prolonged temporal oscillation of physical quantities, such as oscillating congested traffic (OST) and triggered stop-and-go waves (TSG). In the literature, such phenomena were also studied and reproduced by other studies \cite{traffic-flow-hydrodynamics-04,traffic-flow-hydrodynamics-07,traffic-flow-hydrodynamics-15,traffic-flow-hydrodynamics-13,traffic-flow-hydrodynamics-14,traffic-flow-hydrodynamics-12}. 
In fact, the periodic orbit in dynamical system is a topic of increasing interest in various areas of science, known as {\it limit cycle} \cite{ordinary-differential-equation-teschl}.
In the case of traffic flow, they are understood by most authors as distinguished traffic states \cite{traffic-flow-phenomenology-09}. 
It is worth noting that in the context of traffic flow, the physical content of the above oscillatory phases is usually attributed to the inhomogeneity of the traffic system \cite{traffic-flow-hydrodynamics-15,traffic-flow-hydrodynamics-12} and are treated by introducing either some extra external source (such as on-ramp flow) into the system, or local discontinuity into the model parameters.
As far as the authors are aware, in the literature most studies were for specific traffic scenarios with appropriately chosen parameters as well as initial/boundary conditions. 
Besides, as most results were drawn based on numerical simulations with properly chosen initial and/or boundary conditions, the properties and interpretation of the resulting traffic state sometimes vary in different models.
In addition, although the temporal oscillatory traffic state is generally recognized as a traffic phase, its connection to the concept of phase transition in thermodynamics is somewhat obscure.
This motivated us to elaborate the starting point of our approach of the traffic system from the general concepts of stability and of phase transition, rather than from any particular characteristic of the traffic system.
The non-stationary nature connected to the time dependence of measured physical quantities seemingly rules out the possibility to employ a theoretical framework which only deals with quasi-static process and/or steady state, such as the theory of phase transition in thermodynamics. 
Nonetheless, it remains unclear and therefore captivating whether such oscillatory states can also be understood by an approach following the very spirit of the potential function.

The present work aims at a dynamical traffic theory, which reproduces the main characteristic of observed traffic flow data but expressed in the language of the catastrophe model. 
In particular, we study stationary as well as non-stationary traffic phases and the transitions among these states by investigating the stability of the corresponding EoM of the traffic system. In our model, the EoM is based on the kinetic theory of gas, namely, a mesoscopic Boltzmann-type theory. It is shown that the potential function can be defined in terms of the transition coefficients of the Boltzmann-type equation. For the case of steady state traffic, the process to minimize the potential function is equivalent to pick out the stable steady states of the EoM. In this context, our approach ``connects the dots" between the catastrophe model and Landau's theory of phase transition from the viewpoint of a mesoscopic dynamical theory. Moreover, our theory explicitly depends on time, which makes it feasible to investigate the temporal evolution of system and the possibility of any (stable) oscillatory solution.
For the most part, our study strives to establish a general framework of a dynamical theory rather than to focus on specific parameterization.
The paper is organized as follows. In section II, we give a brief review on the traffic flow theories, and then we present the mesoscopic traffic flow model. 
In section III, we investigate steady state traffic, the properties of instability of the EoM, as well as realistic physical constraints. 
We show that the present model possesses most features of the catastrophe model once the {\it potential function} is properly defined in terms of the transition coefficients. Furthermore, the potential function of the model can be seen on the same footing with the {\it thermodynamical potential} and the relation to the Landau's theory of phase transition is established. The model in one-dimensional configuration space is investigated. In section IV, in order to study non-stationary states, we generalize the model to the higher-dimensional case. It is shown that such complication provides the possibility to model the observed temporal oscillation of traffic flow states. We investigate two possible scenarios: in the first scenario, the observed oscillation in traffic flow data is understood to be related to the limit circle solution in configuration space; in the second scenario, the potential function is parameterized to possess a metastable region in accordance to the definition of the {\it synchronized flow} introduced in the three phase traffic theory \cite{traffic-flow-three-phase-01,traffic-flow-review-01}. In both cases, the temporal oscillation is attributed to the curl of a vector potential.
The fundamental diagram in both cases are also studied. The last section is devoted to discussions and concluding remarks.

\section{II. A mesoscopic traffic flow model}

One important empirical measurement for a long homogeneous freeway system is the so called {\it fundamental diagram} of traffic flow. It is plotted in terms of vehicle flow as a function of vehicle density (or concentration). The flow-concentration relation possesses the following main characteristics:
(1) It is divided into two different regions of lower and higher vehicle density respectively, which correspond to {\it free} and {\it congested} flow;
(2) The maximum of the free flow occurs at the discontinuous junction above that of the congested flow region which forms an inverse-$\lambda$ shape and
(3) Congested flow presents a broader scattering of the data points on the flow-concentration plane. 
Recently, we proposed a mesoscopic model for the traffic flow \cite{traffic-flow-btz-lob-01} by the method of stochastic differential equation (SDE) \cite{stochastic-difeq-oksendal}, where the dynamics of the system is governed by the temporal evolution of the distributions of vehicles among different speed states via transitions in configuration space.
In addition to the conventional transition terms, stochastic transition is introduced in the model to describe the stochastic nature of traffic flow.
It was shown that analytic solutions can be obtained not only for the expected value of speed and traffic flow, but also for their variances. 
The latter is understood to account for the resulting scatter of the data point of the fundamental diagram.
In particular, we show that a simplified version of our model by assuming only two
speed states with constant transition coefficients is able to reproduce the empirical data on a Brazilian highway \cite{traffic-flow-data-02}, and two of the above characteristics, (1) and (3), were well-reproduced.

In the following, we briefly review our traffic model, more details of the model can be found in \cite{traffic-flow-btz-lob-01}.
Let us consider a section of highway where the spatial variation of the vehicle distribution can be ignored. For simplicity, only discrete values for speed, namely $v_1,v_2,\cdots,v_D$ are considered, and the number density of vehicles conducting at speed state $v_i$ is denoted by $n_i$. In time, a vehicle with speed $v_i$ may transit to another state $v_j$ according to the following set of SDE 
\begin{eqnarray}
\frac{dn_i}{dt} = \sum_{j=1}^D c_{ij} n_j + \sum_{j=1}^D s_{ij} \sqrt{n_j}w_j
\label{bten}
\end{eqnarray}
where the speed transition on the r.h.s. of the equation is a summation of two contributions: the deterministic and stochastic transitions measured by the transition rates $c_{ij}$ and $s_{ij}$ respectively.
For the stochastic transition, randomness is introduced through the white noise $w_j$, which is a random signal characterized by a featureless power spectral density. 
When $j\ne i$, the coefficients $c_{ij}$ and $s_{ij}$ measure the rate for a vehicle with speed $v_j$ to transit to another state with speed $v_i$.
It has been shown \cite{traffic-flow-btz-lob-01} that these transition coefficients are not completely independent, and the number of degrees of freedom of the system is $D-1$.

It is noted when one is only interested in the temporal evolution of the {\it expected value} of physical quantities, as in many cases of physical interest, the stochastic terms can be ignored \cite{stochastic-difeq-oksendal}. 
By removing the stochastic transition terms, Eq.(\ref{bten}) reduces to the corresponding deterministic EoM
\begin{eqnarray}
\frac{dn_i}{dt} = \sum_{j=1}^D c_{ij} n_j 
\label{btendet}
\end{eqnarray}
As discussed above, different traffic flow phases are originated from the properties of stability of Eq.(\ref{bten}).
For the present model, in principle, there are at least two sources of instability. 
The first one is due to the stability of the deterministic EoM Eq.(\ref{btendet}), 
which is completely determined by the deterministic transition rate $c_{ij}$.
The second source is due to the stochastic fluctuation, measured by the white noise $w_j$ and its strength in terms of the transition rate $s_{ij}$.
In this work, we will concentrate on the first source. The effect of the stochastic transitions was studied in \cite{traffic-flow-btz-lob-01}.

To discuss the stability of the deterministic EoM Eq.(\ref{btendet}), one can introduce a small perturbation around a steady solution.
In general, the transition coefficients are not expected to be constants.
Nevertheless, for a linear system, they usually can be expanded around those of the steady solution. 
Intuitively, for any physical solution, $n_i$ must be bounded from above and below. 
As a result, a stable steady solution implies that the transition matrix, the $(D-1)\times(D-1)$ matrix defined by the transition coefficients, must be locally negative definite.
This is because, any positive eigenvalue would imply that the vehicle number of some state increases exponentially in time near the vicinity of the steady solution, which amplifies any small perturbation. 
On the other hand, for a realistic traffic model which provides a natural description of traffic congestion, 
its EoM must contain region with unstable solutions.
This is the main topic to be explored in the subsequent sections.

\section{III. Potential Function in a one-dimensional Boltzmann-type traffic flow theory}

In this section, let us first concentrate on the simplest case by employing a simplified version of the model which contains only two speed states:
$v_1$ and $v_2$ with corresponding occupation densities $n_1$ and $n_2$.
Using this simple model, our goal is to elaborate, to the greatest extent possible, a traffic flow theory which consists of different traffic phases 
and the transition between them based on the properties of 
the instability of the corresponding EoM.
The equation of motion of this simplified model reads
\begin{eqnarray}
\frac{dn_1}{dt} &=& -p_{11}n_1 + p_{12}n_2 \nonumber \\
\frac{dn_2}{dt} &=& -p_{22}n_2 + p_{21}n_1
\label{bte2}
\end{eqnarray}
where one assumes $v_1 < v_2$ without losing generality. 
When adopting properly chosen parameters \cite{traffic-flow-btz-lob-01}, the above simple model has be shown to reproduce the observed fundamental diagram, as well as its observed variance once the noise terms are present.
By considering the conservation of total vehicle numbers, one has $n_1+n_2=N$, and therefore $p_{11}=-c_{11}=p_{21}=c_{21}$ and $p_{22} =-c_{22}= p_{12}=c_{12}$ ($c_{ij}$ of Eq.(\ref{btendet})). The EoM is in fact of one dimension:
\begin{eqnarray}
\frac{dn_1}{dt} &=& -(p_{11}+p_{12})n_1 + p_{12}N
\end{eqnarray}

For steady solution $n_1^{(0)}$, one has
\begin{eqnarray}
n_1^{(0)} &=& \frac{p_{12}}{p_{11}+p_{12}}N 
\label{steady_condition}
\end{eqnarray}

For a small perturbation around the steady solution, one can expand it as $n_1= n_1^{(0)}+n_1^{(1)}$. When the transition coefficients are simply constants, the resulting EoM of small perturbation reads
\begin{eqnarray}
\frac{dn_1^{(1)}}{dt} &=& -(p_{11}+p_{12})n_1^{(1)}  \label{eomoscill}
\end{eqnarray}
from which one immediately sees that the small perturbations die out exponentially if $p_{11}+p_{12}$ are positive.

However, a model with constant transition coefficients are less physically interesting since they represent a system which is always stable or unstable. In a realistic scenario, a traffic model shall possess different regions featuring distinct stability properties. 
Fortunately, the above argument can be easily generalized to include those elements into the model by considering the the case where transition coefficients are functions of $n_1$.
In particular, it will turn out to be useful to define a {\it potential function}, $U = U(n_1)$, in terms of the transition coefficients by its derivative $U'$ as follows, 

\begin{eqnarray}
U'(n_1) &=& (p_{11}+p_{12})n_1 - p_{12}N  
\end{eqnarray}
In other words, the potential function is determined, up to an irrelevant constant, as follows:

\begin{eqnarray}
U(n_1) &=& \int^{n_1} dn'_1 \left((p_{11}(n'_1)+p_{12}(n'_1))n'_1 - p_{12}(n'_1)N\right) \label{scalarpot}
\end{eqnarray}

\begin{figure}[!htb]
\begin{tabular}{cc}
\begin{minipage}{200pt}
\centerline{\includegraphics*[width=7cm]{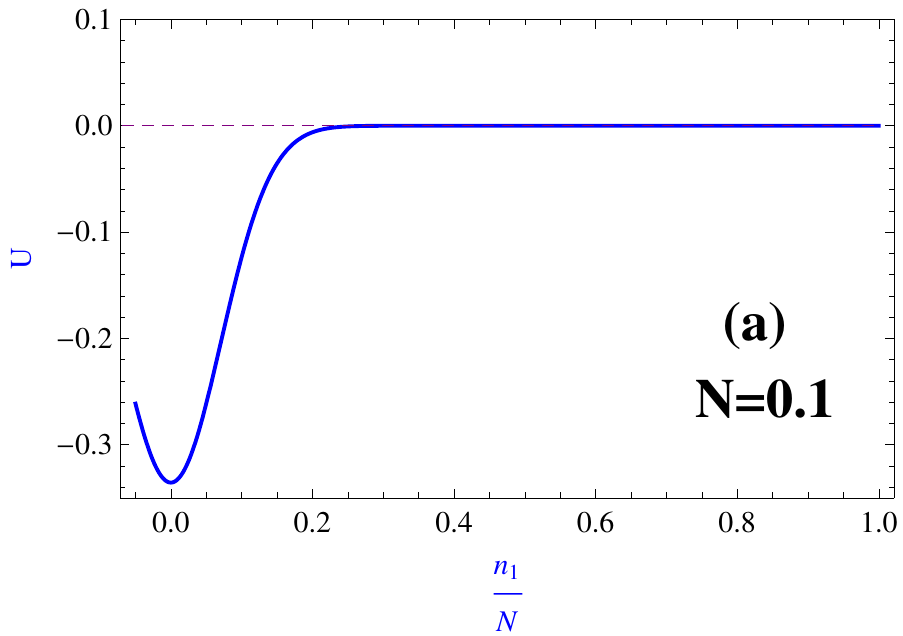} }
\end{minipage}
&
\begin{minipage}{200pt}
\centerline{\includegraphics*[width=7cm]{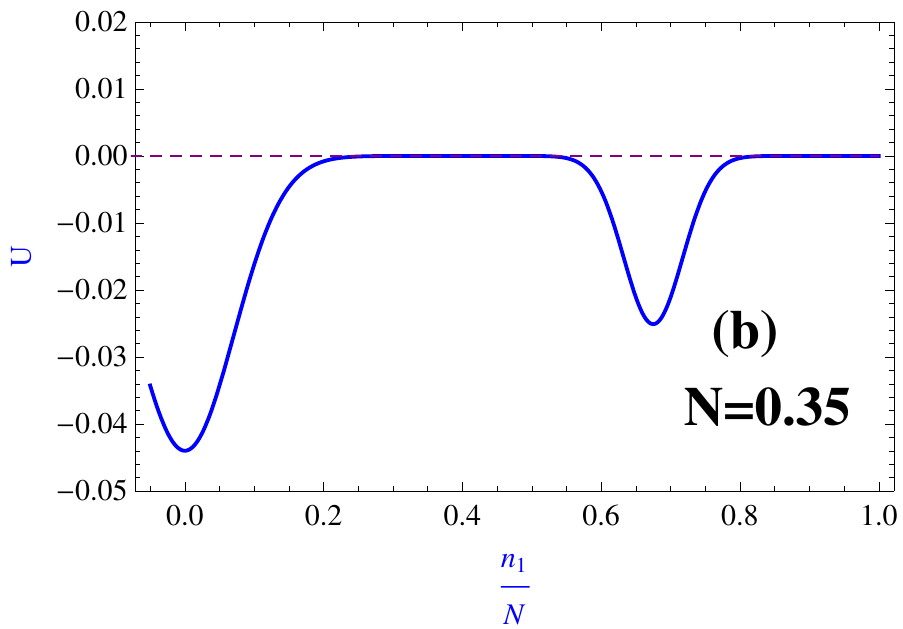} }
\end{minipage}
\\
\begin{minipage}{200pt}
\centerline{\includegraphics*[width=7cm]{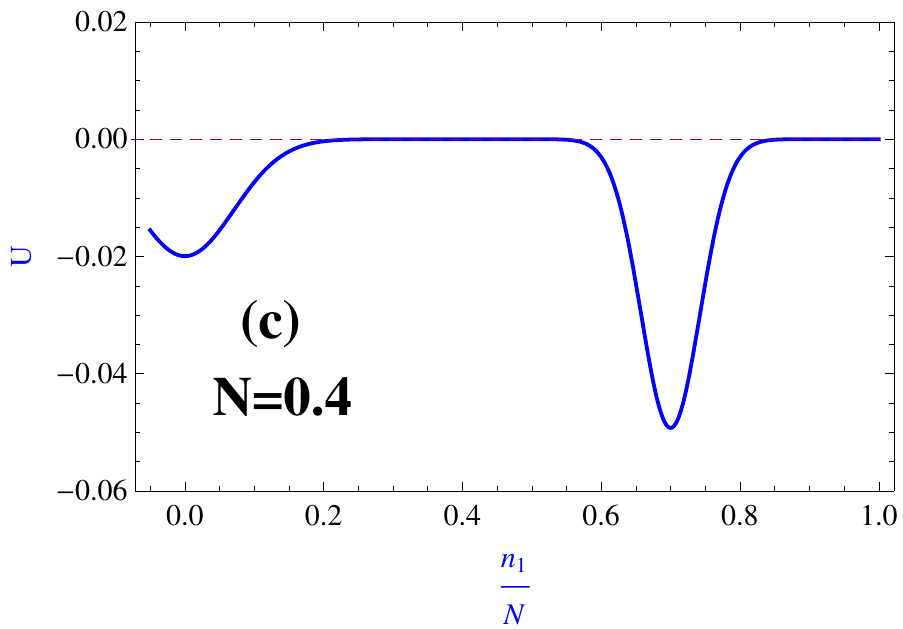} }
\end{minipage}
&
\begin{minipage}{200pt}
\centerline{\includegraphics*[width=7cm]{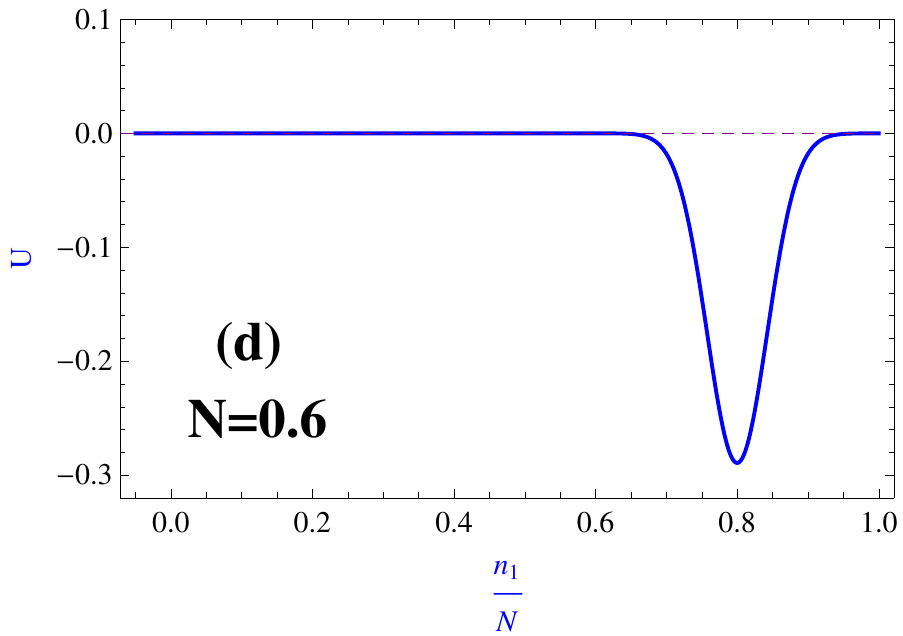} }
\end{minipage}
\\
\end{tabular}
\caption{(Color Online) Schematic potential functions for the one-dimensional case:
\textbf{(a)} Stable free traffic (FT) state at vehicle density $N=0.1$, where the FT state is the only stable steady state;
\textbf{(b)} Stable FT state and metastable homogeneous congestion traffic (HCT) state at $N=0.35$, where both the FT and the HCT states are locally stable solutions, but the FT state is globally stable;
\textbf{(c)} Metastable FT state and stable HCT state at $N=0.4$, similar to the case (b) except that the HCT state is globally stable;
\textbf{(d)} Stable HCT state at $N=0.6$, similar to the case (a), where the HCT state is the only stable steady state.
The analytic expression of the potential function is given in the Appendix.
}
\label{pont1d}
\end{figure}

The potential function defined above is motivated by the fact that for a steady state, one has $\frac{dn_1}{dt}=0$, which implies $U'=0$.
We are about to show that a locally stable state corresponds to a local minimum of the potential function, therefore the potential function defined in Eq.({scalarpot}) has the exactly same physical content as that of the catastrophe model. 
First, it is easy to verify that the equation for small perturbation, Eq.(\ref{eomoscill}), can be generalized and rewritten in terms of the potential function:

\begin{eqnarray}
\frac{dn_1^{(1)}}{dt} &=& -U'' n_1^{(1)}  \label{eomoscill2}
\end{eqnarray}
where one has $U''=(p_{11}+p_{12}) - p_{12}'N + \frac{(p_{11}'+p_{12}')p_{12}}{(p_{11}+p_{12})}N$, the derivatives are carried out with respect to $n_1$ and higher order terms are ignored.
From Eq.(\ref{eomoscill2}), one sees that small perturbation dies out when $U$ is convex around the steady state. It implies that a stable steady state corresponds to a point with $U'=0$ and $U''>0$, namely, a local minimum of the potential function.
Now we are in a position to study the implications of the above potential function on traffic flow theory. 
Since one expects that the model have different stable states for different traffic scenarios, it implies: (1) when the overall vehicle density is small, FT state must correspond to a stable state; (2) when the overall vehicle density is large, the system is likely to be ``trapped" in a HCT state.
Different traffic scenarios discussed above imply that the potential function can not be fixed but it should depend on some quantities that characterize the system, such as the overall vehicle density.
In addition, at very low vehicle density, FT must be a global as well as unique stable state, as one never spots a traffic jam at mid-night. 
And the same arguments are also valid for the stability of the HCT state at very high vehicle density.
When a steady state is globally stable, it means that there is a unique local minimum in the corresponding potential function $U$. 
The potential functions in these cases are illustrated in Fig.\ref{pont1d}(a) and (d) with different total vehicle density $N$, where $N$ is normalized to a range between 0 to 1.
Now we consider an intermediate vehicle density, where a steady FT state characterized by bigger traffic flow and a HCT state characterized by smaller traffic flow coexist. 
In this scenario, both states are locally stable, but only one of them is globally stable. 
In other words, the system locates in a local minimum of the potential function instead of the global minimum, namely, a metastable state.
It is stable against small perturbation until the magnitude of the perturbation exceeds a certain critical value, which is determined by the depth of the local minimum.
This scenario is in fact well-known and discussed by many authors in terms of different traffic models \cite{traffic-flow-hydrodynamics-13,traffic-flow-micro-17}.
In our approach, it is naturally implemented in terms of the potential function, which is shown illustratively in Fig1.(b) and (c).
We note that the coexistence of metastable and stable traffic states brings up an intuitive explanation of the discontinuity presented in the fundamental diagram as shown below.
It is also not difficult to see that the above discussions fall in the very same line of arguments as the theory of transition of Landau, the potential function can be viewed to play the same role as the {\it thermodynamical free energy}. 

Moreover, one has to take into account the physical constraints as boundary conditions.
Since the potential function $U$ is defined to be a function of $n_1$, obviously the system shall only evolve inside the region where $0<n_1< N$. The latter implies that $\frac{d n_1}{dt} \ge 0$ when $n_1 \rightarrow 0$ and $\frac{d n_1}{dt} \le 0$ when $n_1 \rightarrow N$, which consequently apply restrictions on the shape of the potential function: $U'<0$ when $n_1 \rightarrow 0$ and $U'>0$ when $n_1 \rightarrow N$.
In the one-dimensional case, this means that the potential function must be confined between its boundaries at $n_1 =0$ and $n_1=N$.
In the case when one only considers two traffic flow phases, the shape of the potential function is unambiguously determined and it shall look like those depicted in Fig.\ref{pont1d}.

\begin{figure}[!htb]
\begin{tabular}{cc}
\begin{minipage}{200pt}
\centerline{\includegraphics*[width=7cm]{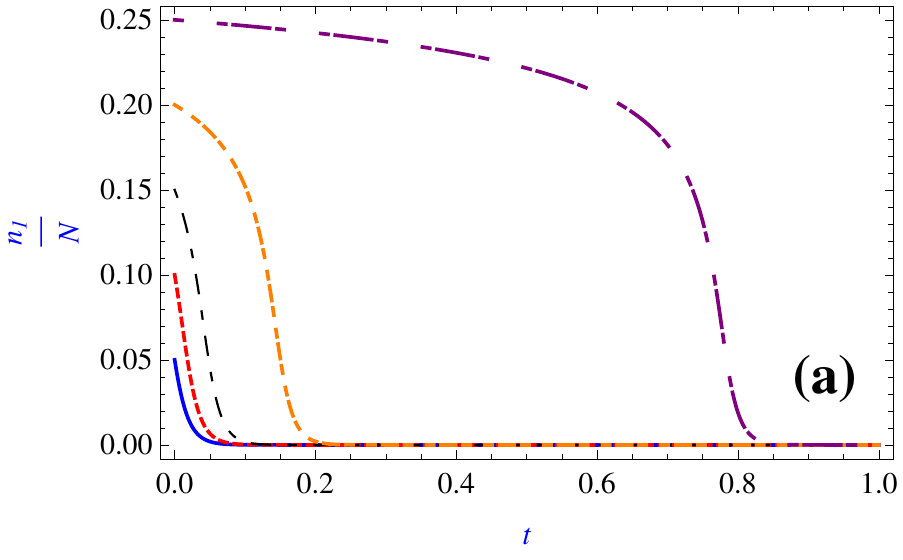} }
\end{minipage}
&
\begin{minipage}{200pt}
\centerline{\includegraphics*[width=7cm]{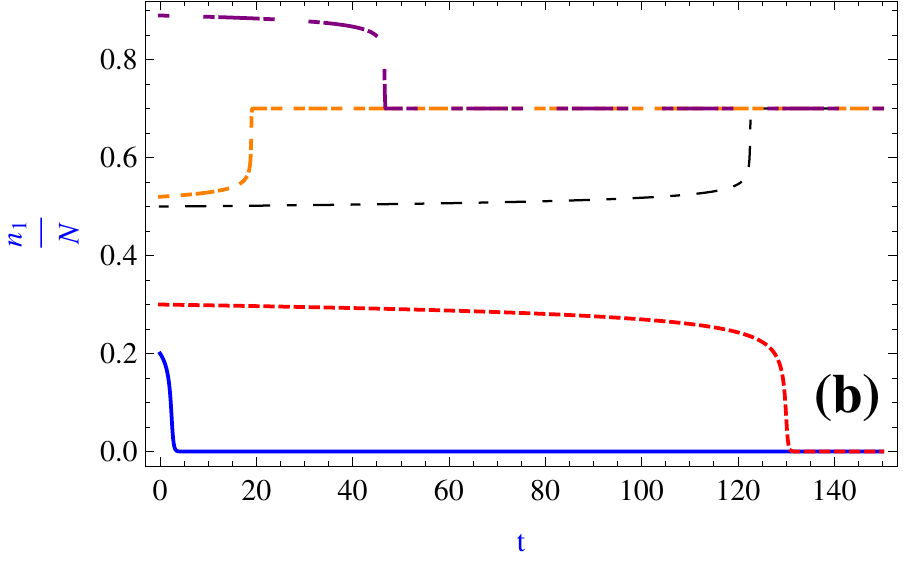} }
\end{minipage}
\\
\end{tabular}
\caption{(Color Online) Temporal evolution of the system for different potential functions in the one-dimensional case:
\textbf{(a)} Temporal evolution with different initial conditions for the potential function shown in Fig.\ref{pont1d}(a). It is noted that the system always evolves into the globally stable FT state, independent of the initial condition; 
\textbf{(b)} Temporal evolution with different initial conditions for the potential function shown in Fig.\ref{pont1d}(c). It is noted that the system may evolves into either one of the two locally stable states. This is because either state is locally stable, small perturbation 
with respect to the steady state dies out in time.
}
\label{ev1d}
\end{figure}

As an example, in Fig.\ref{ev1d}(a) and (b), we calculate numerically the temporal evolution of the system with different initial conditions. 
We adopt exactly the potential functions shown in Fig.\ref{pont1d}(a) and (b). 
One observes that in the case of Fig.\ref{ev1d}(a), the system always evolves into the globally stable steady state independent of its initial conditions; in the case of Fig.\ref{ev1d}(b), the system displays the above mentioned behavior for a metastable state.
The specific form of the potential function used in Fig.\ref{pont1d}-\ref{ev1d} can be found in the Appendix.
By identifying the form of the potential function, it is made possible to determine the mimimal strength of the pertubation in order to knock the traffic flow from a metastable state into a global state, in other words, the depth of a local minimum of the potential function provides us quantitative information on the criterion of the phase transition.

In comparison to the above model, it is easy to show that the potential function proposed in ref.\cite{traffic-flow-btz-lob-01} is also bounded and dependent on the total vehicle density $N$. It has a unique global minimum, which corresponds to the FT state at low density and HCT state at high density.
However, the location of the minimum changes continuously with $N$, so it does not present any sudden jump, neither the coexistence of different phases.

Generally speaking, when it is necessary, one can introduce more local minima into the potential function to model other local stable states. 
We shall not pursue such complication any further,
but turn to discuss the possibility to obtain non-stationary solution in this one-dimensional model.
At a first thought, one might conjecture that an oscillating state corresponds to some periodic transitions between different states while possibly passing through some unstable local maxima. 
However, it is not difficult to show that the above conjecture is not possible: in a one-dimensional case, the system starts to evolve from any state, the nature of the EoM forces it always to evolve in a given direction in the configuration space without ever turning back, and the evolution will only stop when it encounters a stable state. So unless there is a periodic external perturbation to constantly push the system away from a given stable state, the motion can not be periodic. In our model, since the stochastic noise does not possess any periodicity by definition, even if it were taken into account, it might not have served as the cause of possible periodic solution. 
Therefore one arrives to the conclusion that in the one-dimensional case, there is no periodic oscillatory solution.

\begin{figure}[!htb]
\begin{tabular}{ccc}
\begin{minipage}{150pt}
\centerline{\includegraphics*[width=5cm]{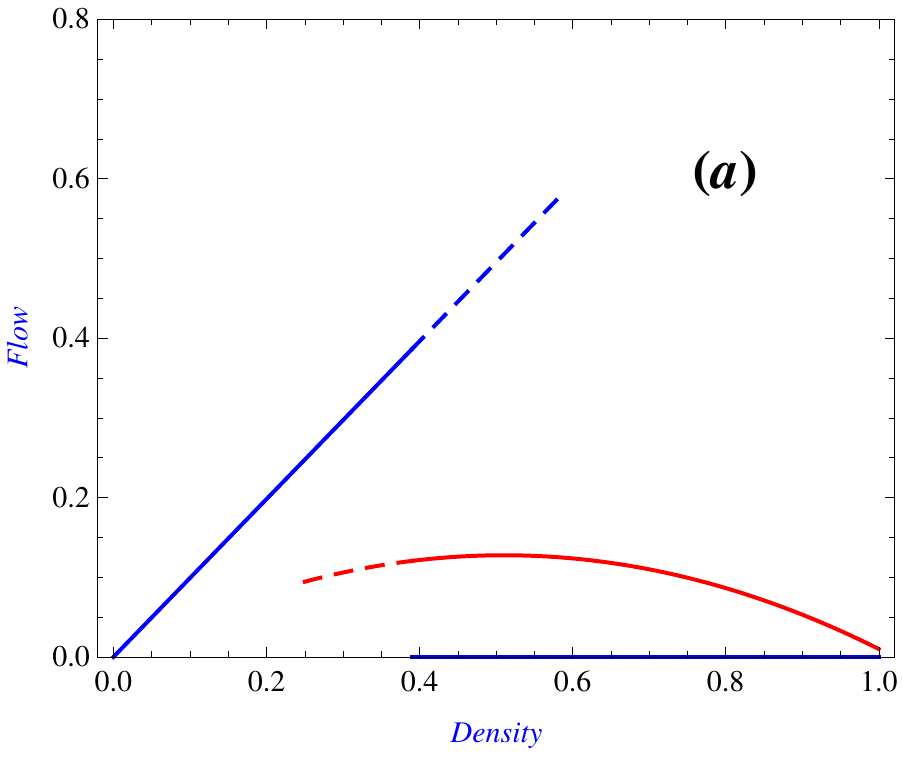} }
\end{minipage}
&
\begin{minipage}{150pt}
\centerline{\includegraphics*[width=5cm]{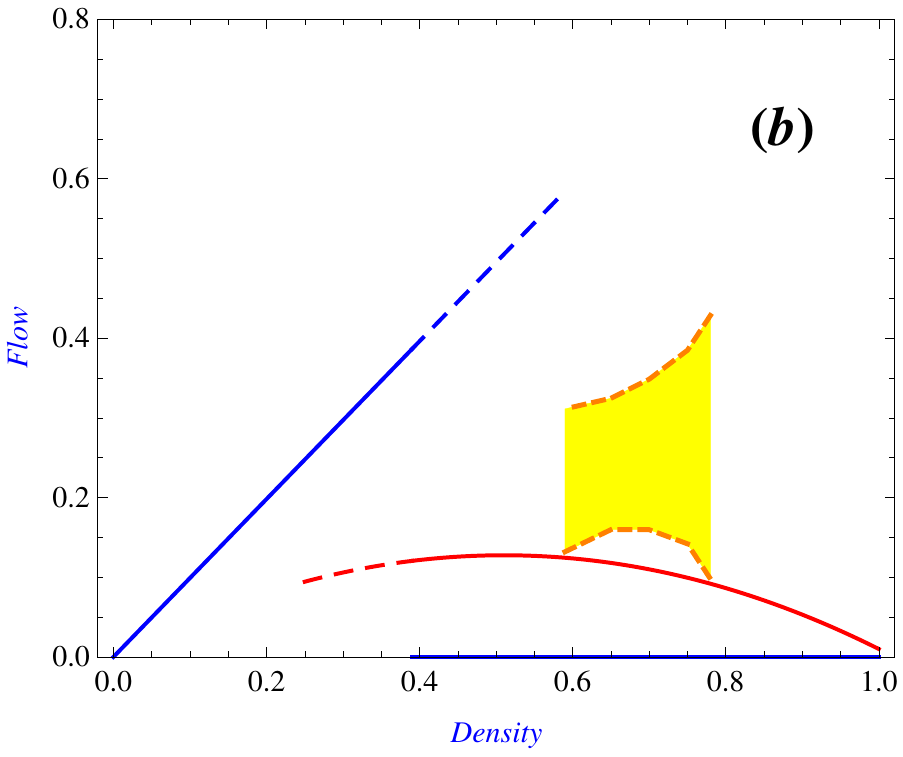} }
\end{minipage}
&
\begin{minipage}{150pt}
\centerline{\includegraphics*[width=5cm]{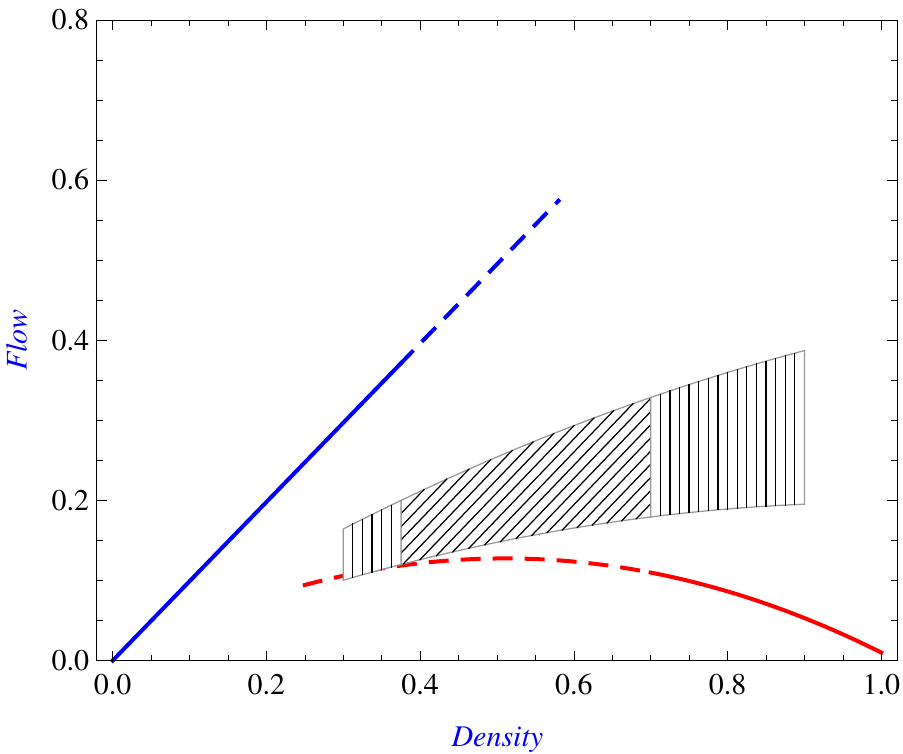} }
\end{minipage}
\\
\end{tabular}
\caption{(Color Online) Fundamental diagram of the present traffic model:
\textbf{(a)} Fundamental diagram of the one-dimensional case where the inverse-$\lambda$ feature is reproduced owing to the jump between the metastable and stable states. Here one uses solid lines to indicate stable states and dashed or dotted lines to indicate metastable states;
\textbf{(b)} Fundamental diagram of the first scenario in the two dimensional case. The scattering of the flow-concentration relationship is due to the stable limit circle solution. When one removes the curl field $\vec{B}=\nabla \times \vec{A}^{(2)}$ and the auxiliary potential $U_2^{(2)}$, the resulting fundamental diagram reduces to that of the the inverse-$\lambda$ shown in (a);
\textbf{(c)} Fundamental diagram of the second scenario in the two dimensional case. The scattering of the flow-concentration relationship is due to a finite region of metastable states. Again, if one removes those metastable states, the resulting fundamental diagram reduces to that of the one-dimensional case shown in (a).
}
\label{fdiagram}
\end{figure}

In what follows, we calculate the fundamental diagram obtained from this one-dimensional model, and present it in Fig.\ref{fdiagram}(a). 
In our calculation, for simplicity, we assume $v_1=0, v_2=1$. The details of the formulae can be found in the Appendix. 
The calculated fundamental diagram shows clearly an inverse-$\lambda$ shape. 
This is owing to the coexistence of the metastable states (dashed or dotted curves) and the stable states (solid curve), as discussed in the text concerning Fig.\ref{ev1d}. 
The maximum flow of the FT state (the tip of the inverse-$\lambda$) corresponds to a metastable FT state, while the corresponding stable steady solution is represented by a stable HCT state lying vertically below it: perturbations big enough may trigger the system to jump from the tip of the inverse-$\lambda$ to the HCT state.
The jump is due to the difference of the magnitudes of the corresponding traffic flow of the two states, which is absent in the model proposed in \cite{traffic-flow-btz-lob-01}. 
However, since one is only considering the deterministic EoM, the current approach does not contain a mechanism within itself to provide such perturbation.
We note that such {\it sudden jump} of physical quantities appears naturally in our approach, in the sense that it is not originated from any discontinuity in the parametrization of the model, but from its nonlinearity. 
However, another important feature of the fundamental diagram, namely, the observed scattering of the data of the flow-concentration relation in the congestion region (as shown in Fig.\ref{fdiagram}(b) and Fig.\ref{fdiagram}(c)), is not reproduced here. Intuitively, such feature may partly be related to the stochastic nature of the traffic flow \cite{traffic-flow-btz-lob-01}. Interestingly, as discussed below, it may also be attributed to either a stable limit circle solution or to a finite range of metastable solution in the parameter space. However, in the one-dimensional case, neither of the two solutions is attainable, they only exist in the higher dimensional case, as to be discussed in the following section.

\begin{figure}[!htb]
\begin{tabular}{c}
\begin{minipage}{200pt}
\centerline{\includegraphics*[width=7cm]{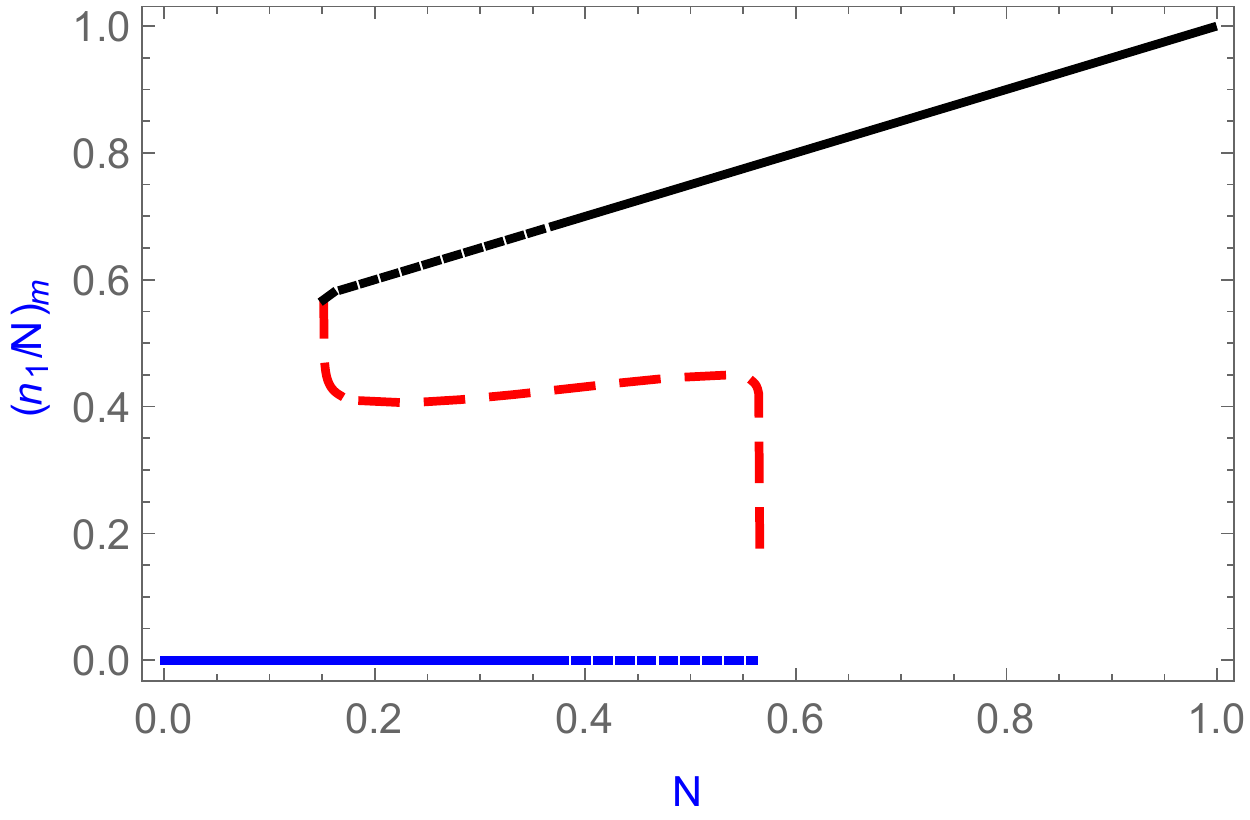} }
\end{minipage}
\\
\end{tabular}
\caption{(Color Online) Extrema of the corresponding ``cusp catastrophe" model in its parameter space. Solid line corresponds to global minima, while dashed and dotted lines indicate local maxima and minima respectively. 
For instance, at $N=0.35$, the extrema is threefold, its potential function is shown in Fig.\ref{pont1d}(b); 
while at $N=0.6$, the extremum is onefold, its potential function is shown in Fig.\ref{pont1d}(d). 
}
\label{extrema_catas_d1}
\end{figure}

To close this section, we give a few more comments on the connection between the ``cusp catastrophe" model and the Landau's theory of phase transition.
According to Landau \cite{statistical-mechanics-landau}, the stability of a thermodynamic system at fixed temperature and volume is governed by its free energy, which can be derived from the partition function following standard procedure of statistical physics.
In this phenomenological approach, the free energy of the system can be expressed as a function of order parameter $\Psi$.
A metastable state corresponds to a local minimum of the free energy, while a macroscopically stable state is encountered at the global minimum of the free energy.
If the free energy has only one minimum, the thermodynamical system possesses only one phase.
At the moment when a second phase begins to emerge, the extrema of the free energy become threefold, namely, two minima and one maximum.
A phase transition takes place when the free energies of the two phases are equal.
The phase transition is said to be of second order if the two phases coincide at a critical point, otherwise the phase transition is of first order.
In the catastrophe model, it is assumed that the physical state is attained at the global minimum of the potential function.
Apart from the fact the potential function is usually an a priori ingredient of the model, the mathematics of the two approaches are strikingly similar.
In the case of ``cusp catastrophe" model, the potential function is threefold in some specific region of the parameter space and onefold otherwise.
To illustrate the above discussions clearly, the extrema of the above one dimensional model is shown in Fig.\ref{extrema_catas_d1}, 
where the positions of the extrema are determined by steady states of the equation of motion Eq.(\ref{steady_condition}).
At $N=0.35$ for instance, two of the three extrema correspond to stable solutions while the other one is linear unstable.
In our model, the potential function is derived in terms of transition coefficients by Eq.(\ref{scalarpot}), therefore it provides a mesoscopic interpretation for the catastrophe model stemmed from the dynamics of the traffic system.

\section{IV. Higher dimensional case and temporal oscillatory solution}

The above discussions on scalar potential function can be generalized to the higher-dimensional case.
The motivation to study the problem in higher dimensional parameter space is to investigate the non-stationary solution of traffic system.
To increase the dimension of the parameter space,
one notes that the set of equations Eqs.(\ref{bte2}) introduced above can equivalently be seen as a system with two degrees of freedom if no constraint is applied. Following the same arguments of the previous section, 
one can write down the EoM in terms of the gradient of a potential function $U$ and define it as a vector field $\vec{E}$ as follows
\begin{eqnarray}
\frac{dn_1}{dt} =  E_1 \label{eq2d1}  \\
\frac{dn_2}{dt} =  E_2 \nonumber \\
\vec{E} = (E_1,E_2) \nonumber 
\end{eqnarray}
with the vector field $\vec{E}$ defined as the gradient of the potential:
\begin{eqnarray}
\vec{E} &=& - \nabla U
\label{poteq1}
\end{eqnarray}

\begin{figure}[!htb]
\begin{tabular}{ccc}
\begin{minipage}{130pt}
\centerline{\includegraphics*[width=4.5cm]{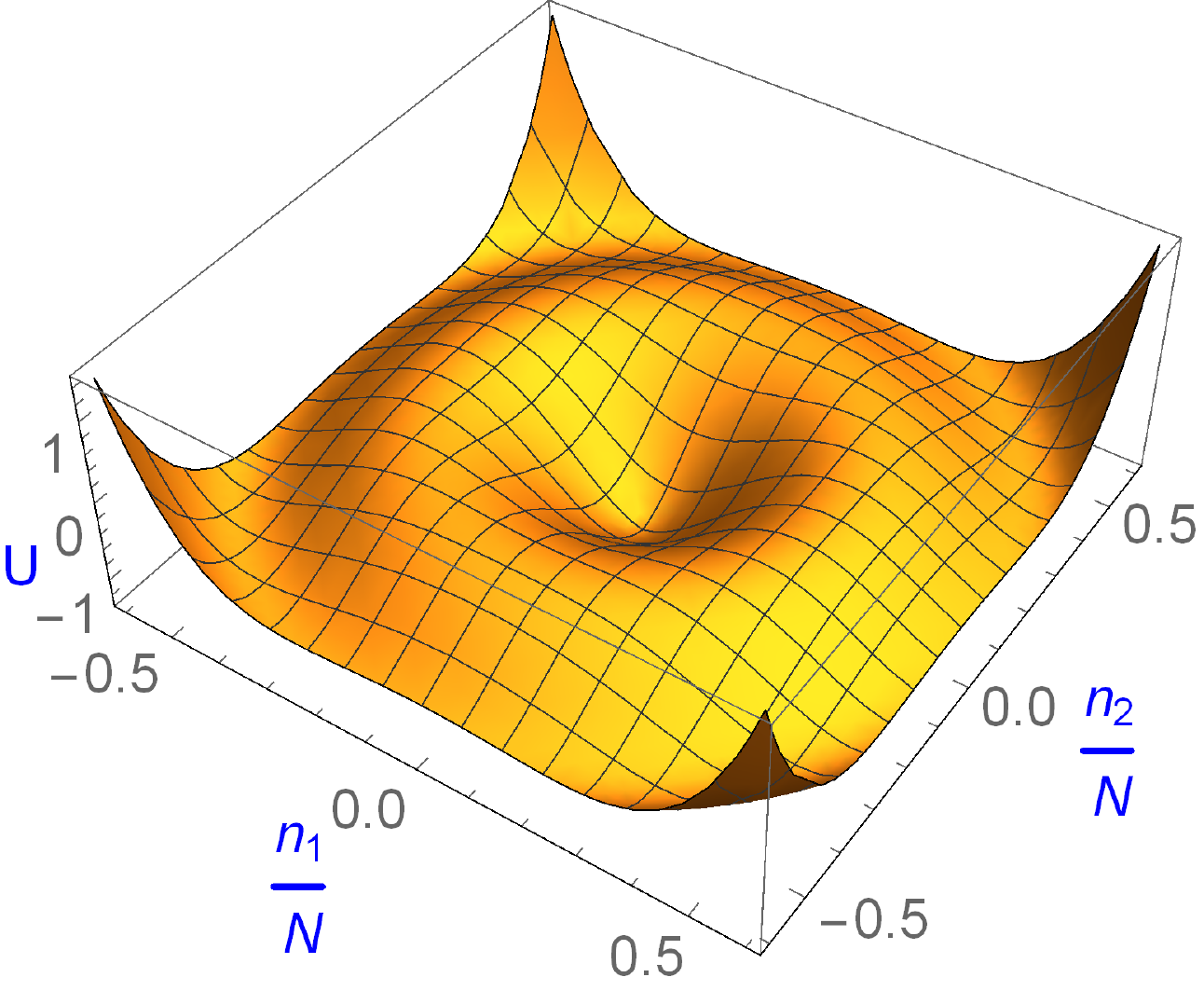} }
\end{minipage}
&
\begin{minipage}{130pt}
\centerline{\includegraphics*[width=4.5cm]{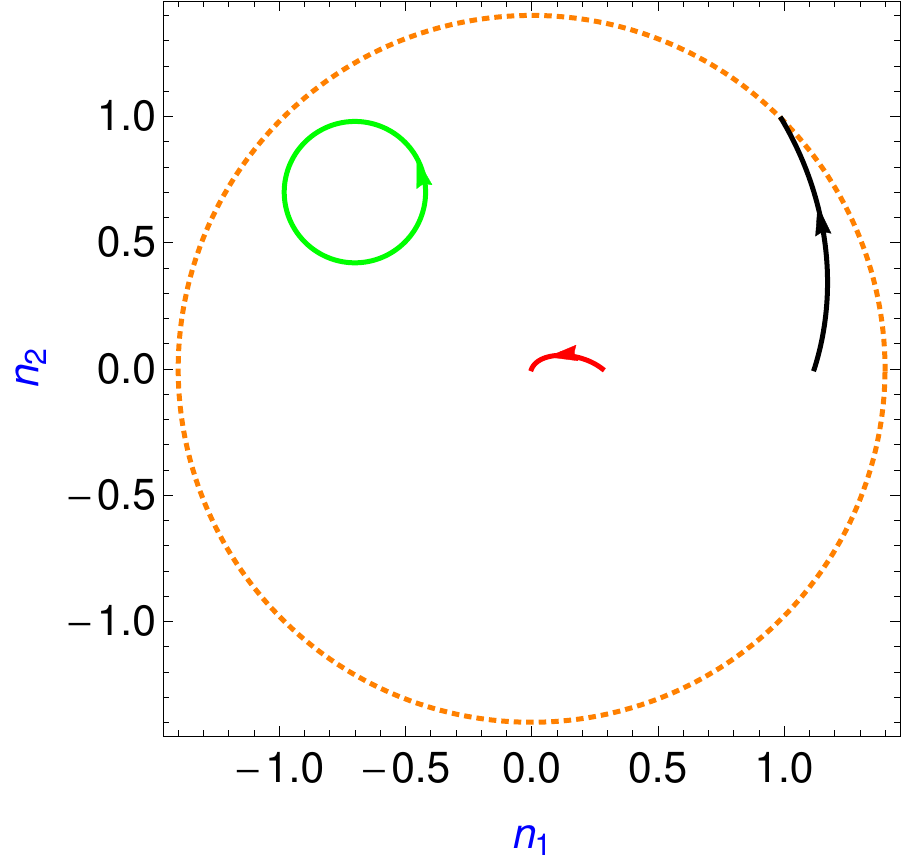} }
\end{minipage}
&
\begin{minipage}{130pt}
\centerline{\includegraphics*[width=4.5cm]{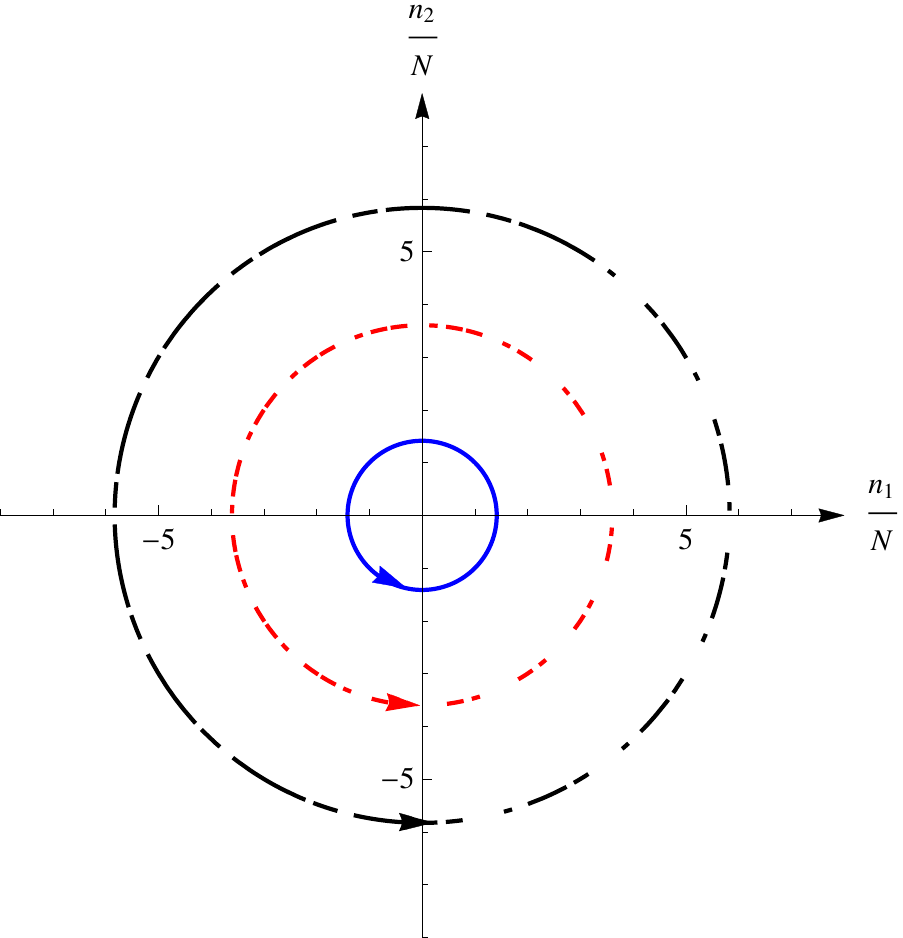} }
\end{minipage}
\\
\end{tabular}
\caption{(Color Online) Schematic potential function of the two-dimensional case and possible solutions:
Left: An illustrative two-dimensional potential featured by more than one local minimum: the minimum at the center of the potential represents the FT state, other minima correspond congestion states; Middle: Schematic temporal evolution in the configuration space. The red and black curves indicate the temporal evolution of the system against small deviations from FT and congestion states respectively, in both cases, the system quickly converges to the nearest stable state. The green circle indicates a possible scenario of a limit circle solution in the two-dimensional parameter space;
Right: Semi-stable limit circle solutions of Eqs.(\ref{eq2d2}).
}
\label{pont2d1}
\end{figure}

It is noted that a given state of the system corresponds to one point in the configuration space; the system evolution is completely determined by the vector field alone, which draws a curve starting from the point in the configuration space corresponding to the initial state. 
The evolution of the system can be viewed equivalent as that of a mechanical system according to Hamilton's equation in classical mechanics.
In Fig.\ref{pont2d1}, we depict in the left panel, an illustrative two-dimensional potential and in the middle, how the system can evolve in time in the configuration space. 
The schematic potential has two minima. The first one is at the center of the potential, let us assume that it corresponds to the FT state at low vehicle density; the second minimum locates in the basin that circles around the center, which represents the congestion state; at the outskirt of the potential, the gradients are all pointing inwards, so that the system will not diverge.
Apart from its complication, such a two-dimensional potential does not introduce new physics.
As shown in the middle panel of Fig.\ref{pont2d1}, for small deviations from either FT or congestion state, the system quickly converges to the corresponding stable state.

However, the green circular trajectory in the middle panel of Fig.\ref{pont2d1} indicates a novel possibility.
Unlike the one-dimensional version of the model, now it is possible for the system evolution to trace out a closed curve, therefore to form the so called {\it limit circle} in the parameter space as shown by the circular trajectory in the plot.
Once the system evolves back to its initial condition, the EoM implies that the system will repeat the trajectory and carry out a periodic motion in the parameter space.
Theorem on the existence of the limit circle is described by the Pioncar\'{e}-Bendixson theorem \cite{ordinary-differential-equation-teschl}: roughly speaking, when no stable state exits, the system evolution must be featured by periodic orbit.
Unfortunately, traffic system possesses at least the FT as a stable state, and therefore a general theory on the limit circle of such a system is yet unknown.
In fact, for the specific form of the potential function shown in the left panel of Fig.\ref{pont2d1}, it is easy to show that the above periodic evolution will not occur at all when the vector field is solely determined by the gradient of the potential in Eq.(\ref{poteq1}). 
This is because the potential in Eq.(\ref{poteq1}) is conservative by definition, and therefore the corresponding field is irrotational. 
Nonetheless, an oscillatory solution can be explicitly obtained if one introduces some rotation into the EoM as in the following equations:
\begin{eqnarray}
\frac{dn_1}{dt} = -n_2 \equiv B_1 = -\sqrt{n_1^2+n_2^2}\sin\theta  \nonumber \\
\frac{dn_2}{dt} = n_1 \equiv B_2 = \sqrt{n_1^2+n_2^2}\cos\theta \nonumber \\
\theta = atan(n_2/n_1) \label{eq2d2}
\end{eqnarray}
Since the above vector field is solenoidal (divergenceless), it can be written as follows

\begin{eqnarray}
\vec{B} &=& (B_1,B_2) \nonumber \\
\vec{B} &=& \nabla \times \vec{A}
\label{poteq2}
\end{eqnarray}
where $\vec{A}$ is a vector potential in comparison to the scalar potential $U$ introduced before. 
Due to the fact that the vector field $\vec{B}$ is solenoidal, according to Pioncar\'{e}-Bendixson theorem, it possesses limit circle solution which go round and round in the configuration space perpetually. 
For such a simple case, the corresponding analytic solution is shown in right panel of Fig.\ref{pont2d1} and is known as {\it semi-stable limit circle}. 
In a realistic case, an additional scalar potential may also exist. 
The behavior of the solution depend on the relative strength between the scalar and vector potential.
If one assumes that the scalar potential is strong enough and its stable minimum locates at the center of the circular evolution, any small perturbation may cause the trajectory to eventually sink into the stable minimum. Such solution is known as {\it unstable limit circle}.
Influenced by the shape of and relative strength between the vector field and scalar field, the circular solution may also be stable, so that all nearby trajectories spiral towards it. 
In fact, we understand that most empirically observed periodic oscillations are stable against small perturbations, and therefore must correspond to {\it stable limit circle} as discussed in ref.\cite{traffic-flow-hydrodynamics-13}.
It is observed that one also encounters cases concerning the transition between periodic oscillatory state and different steady stable states.
For instance, by adjusting the vehicle flow at the entrance of the ramp, one may force the traffic phase to transfer from a stable FT into either an OST or a HCT \cite{traffic-flow-hydrodynamics-13,traffic-flow-hydrodynamics-12,traffic-flow-data-09}.
Such phenomena can be interpreted as the transition between stable limit circle and stable critical points.
Additionally, it is known in vector calculus \cite{classical-electrodynamics-jackson} that a general vector field can always be decomposed into a sum of an irrotational field and a solenoidal field, which ensures that the above discussions is generally valid since the EoM can be always written in the following form

\begin{eqnarray}
\frac{dn_1}{dt} = E_1+B_1 \nonumber \\
\frac{dn_2}{dt} = E_2+B_2  \nonumber \\
\label{eq2d3}
\end{eqnarray}
where the vector fields $\vec{E}$ and $\vec{B}$ are determined by the scalar (Eq.(\ref{poteq1})) and vector (Eq.(\ref{poteq2})) potentials respectively.
Roughly speaking, the scalar potential defines the locations of possible sinks, while the vector potential creates possible limit circles, the interplay between the two results in different types of limit circle solutions accompanied by steady solution.

In what follows, we construct two different scenarios, where the model consists of an irrotational field $\vec{E}$ and a solenoidal field $\vec{B}$ in a two-dimensional case. In the first scenario, the model is aiming at reproducing a traffic theory consisting of FT, HCT and OST as studied in \cite{traffic-flow-micro-17}. It is done by properly choosing the form of the scalar and the vector potential to form a stable limit circle for the OST state. 
In the second scenario, we modified our model in order to incorporate the ``synchronized flow" state and the corresponding F$\rightarrow$S transition introduced in the three phase traffic theory. 
There, the vector potential is essential for the oscillatory behavior before the synchronized flow state is attained.

\subsection{The first scenario: an implementation for FT, HCT and OST states}

In this scenario, the potential function consists of a scalar potential and a vector potential.
We relegate the detailed form of the potential functions to the Appendix, 
here we only outlines the rule of thumb for the choice of the potentials: 
(1) The scalar potential has three minima, which give rise to FT, HCT and OST respectively; 
(2) At low/high vehicle density, the minimum of the scalar potential corresponding to the FT/HCT dominates, so that the resulting potential is essentially the same as in the one dimensional case discussed in section III; 
(3) At intermediate vehicle density, an oscillatory solution is found which is caused by the interplay between the scalar and the vector potential where the latter is the strongest in the vicinity of the corresponding minimum of the scalar potential. The latter is tuned to be strong enough to attract the oscillatory solution but not too strong to destroy the stability of the limit circle. 
(4) The range of the above stable limit circle shall reproduce the scattered feature of the observed data.

\begin{figure}[!htb]
\begin{tabular}{ccc}
\begin{minipage}{130pt}
\centerline{\includegraphics*[width=4.5cm]{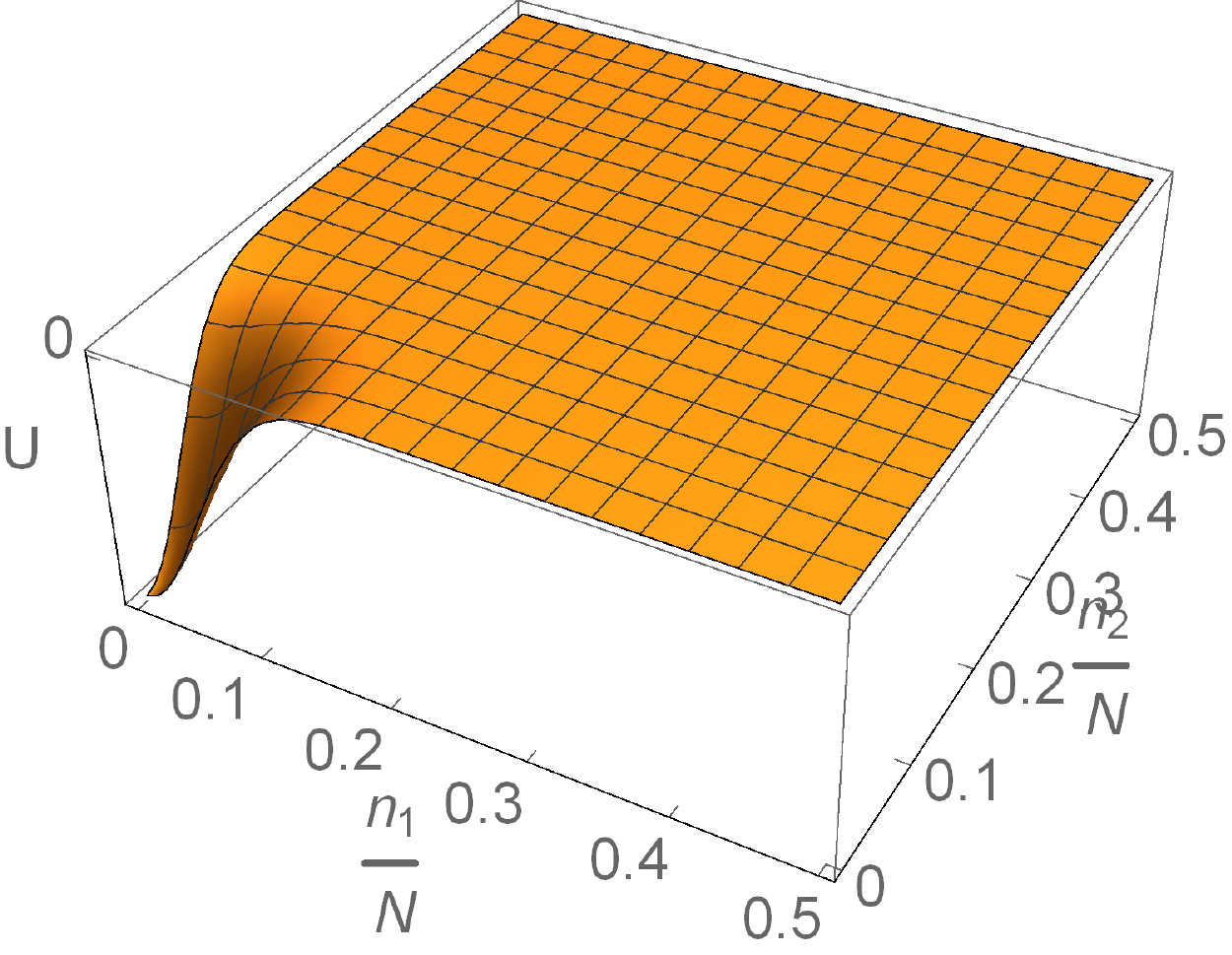} }
\end{minipage}
&
\begin{minipage}{130pt}
\centerline{\includegraphics*[width=4.5cm]{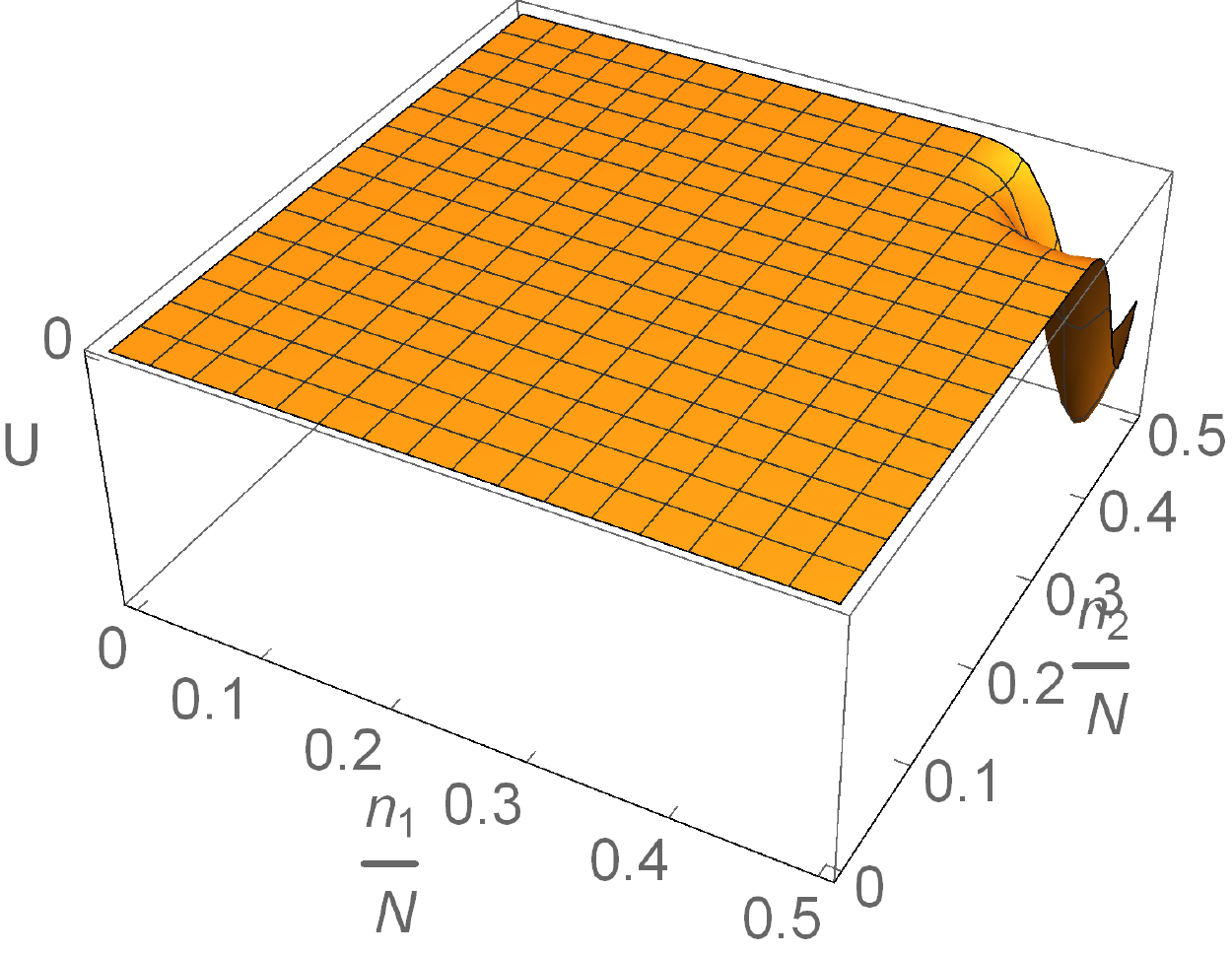} }
\end{minipage}
&
\begin{minipage}{130pt}
\centerline{\includegraphics*[width=4.5cm]{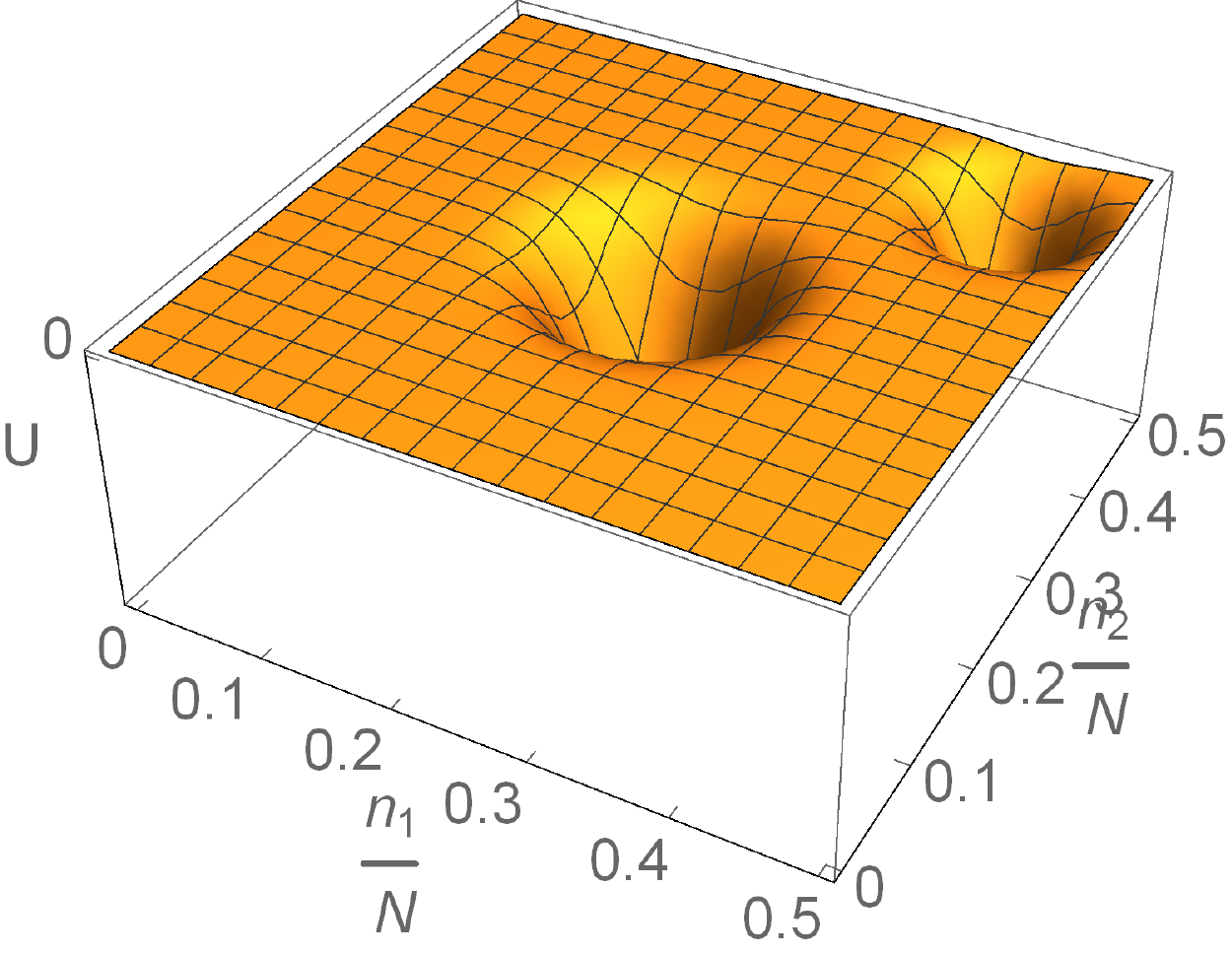} }
\end{minipage}
\\
\end{tabular}
\caption{(Color Online) The two-dimensional scalar potential functions for a three-phase traffic theory: 
Left: The scalar potential at low vehicle density $N=0.2$, it contains the global minimum corresponding to the stable FT state;
Middle: The scalar potential at high vehicle density $N=0.8$, it contains the global minimum corresponding to the stable HCT state;
Right: The scalar potential at intermediate vehicle density $N=0.6$ featuring two local minima, their interplay with the vector field $\vec{B}$ forms a stable limit circle, the OST state. 
The analytic expression of the potential function is given in the Appendix.
}
\label{pont2d2}
\end{figure}

\begin{figure}[!htb]
\begin{tabular}{ccc}
\begin{minipage}{130pt}
\centerline{\includegraphics*[width=4.5cm]{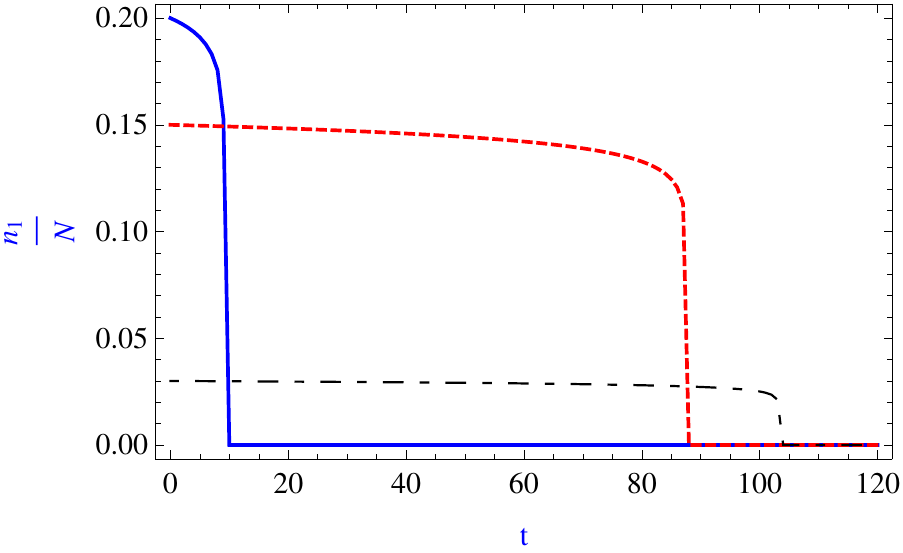} }
\end{minipage}
&
\begin{minipage}{130pt}
\centerline{\includegraphics*[width=4.5cm]{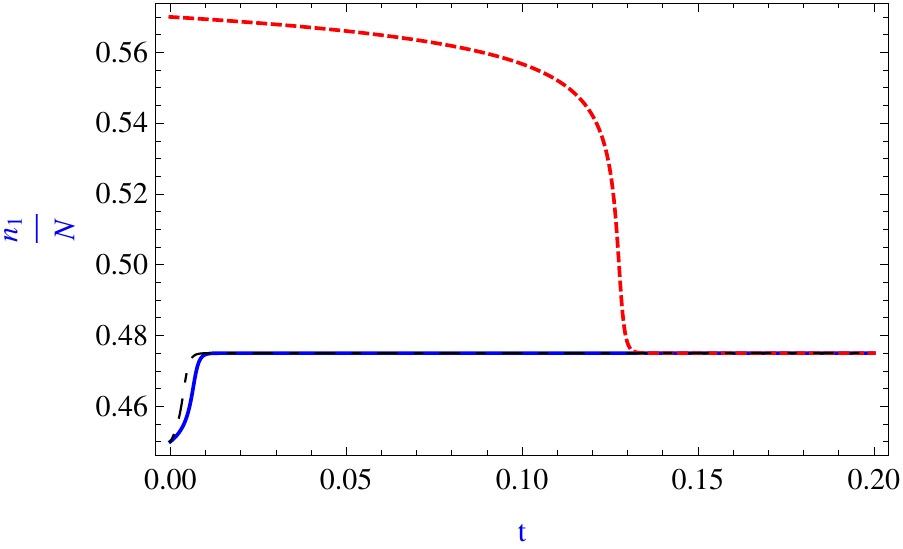} }
\end{minipage}
&
\begin{minipage}{130pt}
\centerline{\includegraphics*[width=4.5cm]{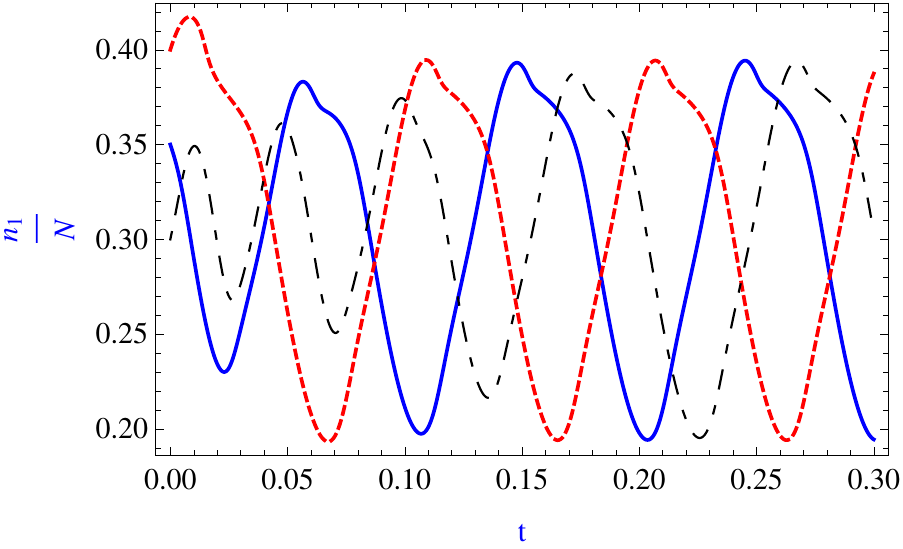} }
\end{minipage}
\\
\begin{minipage}{130pt}
\centerline{\includegraphics*[width=4.5cm]{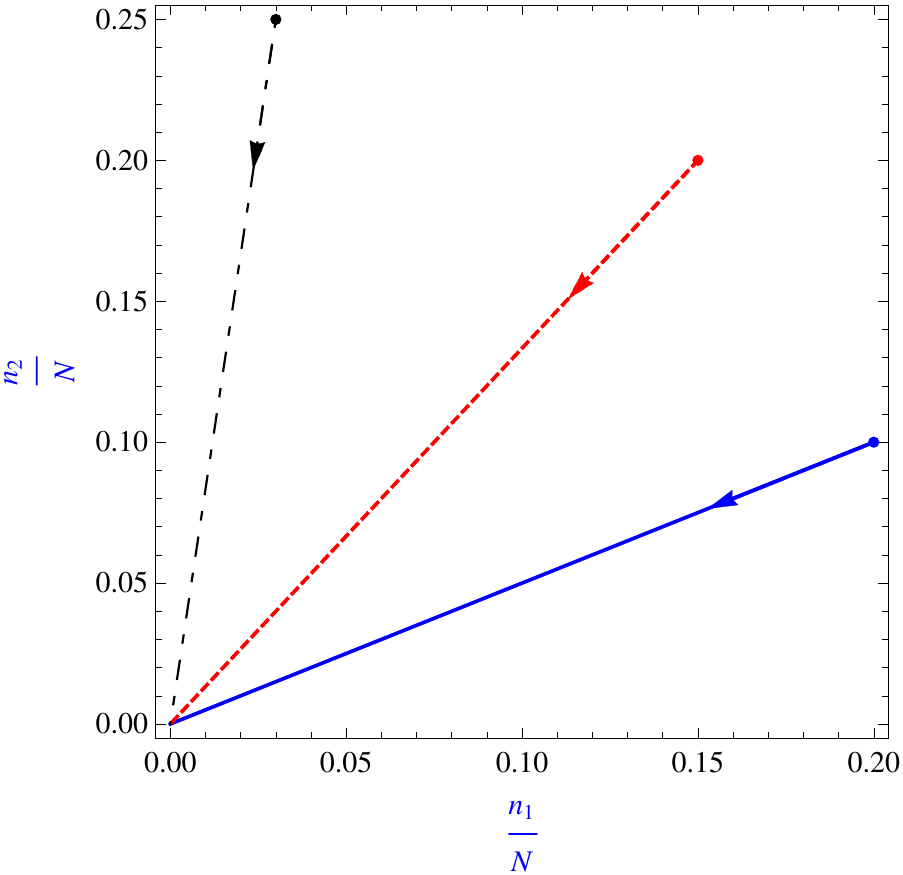} }
\end{minipage}
&
\begin{minipage}{130pt}
\centerline{\includegraphics*[width=4.5cm]{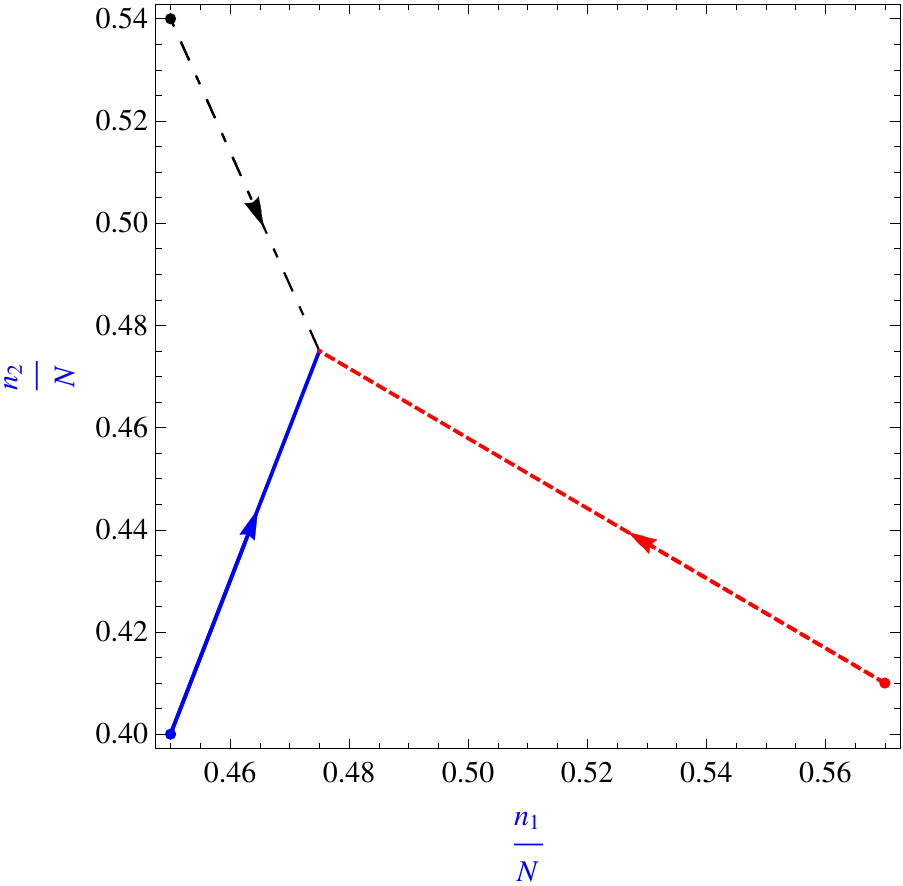} }
\end{minipage}
&
\begin{minipage}{130pt}
\centerline{\includegraphics*[width=4.5cm]{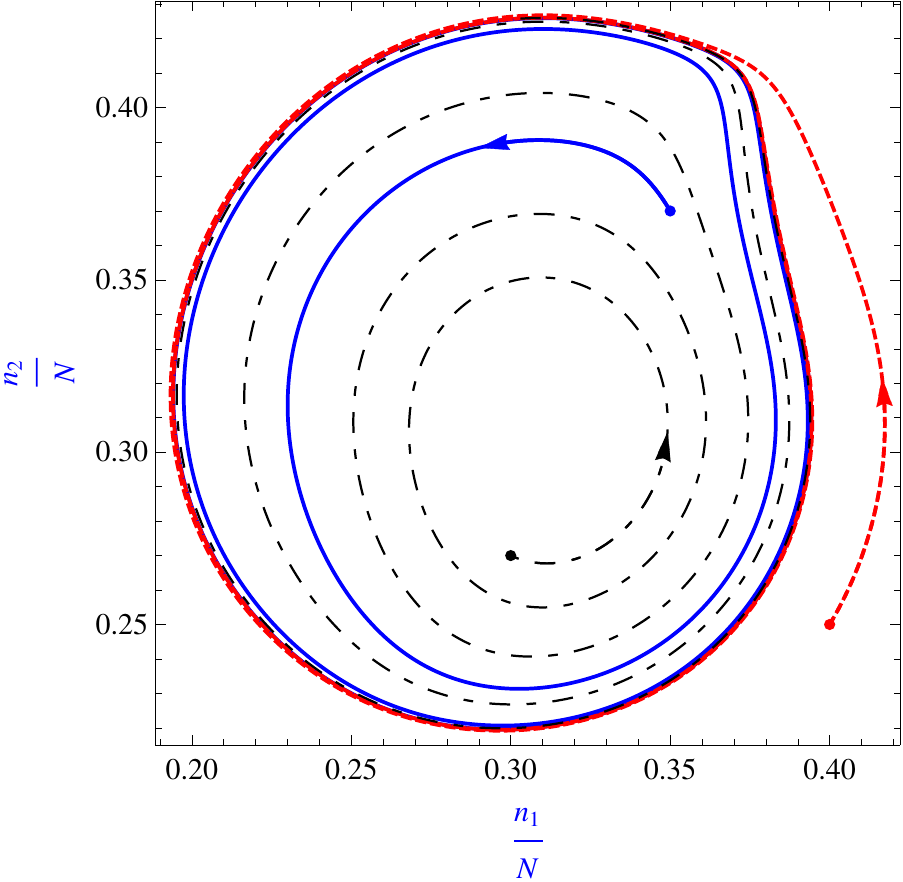} }
\end{minipage}
\\
\end{tabular}
\caption{(Color Online) Temporal evolution of the system for the first scenario of the two-dimensional case.
From the left to the right are the three scenarios shown in Fig.\ref{pont2d2} for low vehicle density, high vehicle density and intermediate vehicle density respectively.
First column: Temporal evolution of one of the parameters of the system. For the FT and the HCT case, the system converges to the corresponding stable state, for the OST case, the system converges to a periodic motion;
Second column: Evolution of the system in the configuration space. For the limit circle solution, we note that the system never converges to any fixed state.
}
\label{ev2d1}
\end{figure}

In Fig.\ref{pont2d2}, we show the scalar potentials of the model for three different traffic states corresponding different vehicle densities.
To show the above model indeed describes the above different traffic states and their transitions, we show the corresponding temporal evolutions of the system in Fig.\ref{ev2d1}.
The results are obtained by numerical calculations. 
To simplify the calculations, we assume that the velocities corresponding to the two free parameters $n_1$ and $n_2$ are almost degenerate, namely, $v_1 \simeq v_2 \simeq 0$, while the third velocity $v_3 = 1$.
The first column of Fig.\ref{ev2d1} depicts one of the parameters, the ratio between the occupation density $n_1$ and the total density $N$, as a function of time.
One notes that other physical quantities (such as overall vehicle density and flow) subsequently follow the same pattern.
In the second column we show how both parameters, $n_1/N$ and $n_2/N$, evolve in the configuration space.
In the plots, different initial conditions are considered.
It is seen that the characteristics of all three phases are well reproduced. 
In the cases of FT and HCT, the system always shows non-oscillatory relaxation to the corresponding stable state.
On the other hand, at intermediate vehicle density, different phases coexist: some initial conditions can involve into OST where the system oscillates and approaches the limit circle, while the others with bigger initial flow are instead drawn into the FT state. 
The oscillatory solution in the third column of Fig.\ref{ev2d1} is identified as OST since the flow oscillates between a value (close to that) of a HCT state and an intermediate value.
It is emphasized since the present model is introduced in a very generic context, the resulting limit circle solution is held in a general context which does not depend on specific choice of parameters.
For instance, TSG can also be easily incorporated into the model if one further introduces another additional rotation involving the minimum of FT state.

The resulting fundamental diagram is shown in Fig.\ref{fdiagram}(b). 
It is pointed out that due to our intended choice of parameters, the curve of FT and HCT states lie exactly at the same place as the one-dimensional version of the model, so that the difference between the two models becomes more transparent.
In the present case, more phases are introduced into the fundamental diagram owing to the complication of the potential functions.
The appearance of the grey band in the fundamental diagram is due to the oscillatory solution of OST, which occupies a finite area in the diagram instead of a thin curve.
Therefore it may account for the scattered feature of the observed data.
It is also noted that stability of FT and HCT phase is modified in some regions, although the positions of the curves remain unchanged.
This is because the existence of a the HCT phase affects the stability of the others. As a result, part of the stable HCT phase (marked in solid red curve) in Fig.\ref{fdiagram}(a) 
turns into metastable phase (marked in dashed red curve) in Fig.\ref{fdiagram}(b); and the part of HCT overlapping with the OST phase simply disappears due to the strong influence of the vector potential.

\subsection{The second scenario: an implementation of synchronized flow state}

For this scenario, the synchronized flow state possesses the following characteristics \cite{traffic-flow-review-01,traffic-flow-review-08}:
(1) According to the three phase traffic theory, the synchronized flow state is related to metastable states.
(2) The collection of metastable states accounts for the scattering data points in a region on the fundamental diagram where the traffic breaks down. As a result, there are infinite number of metastable states in the parameter space. In our model this is implemented by introducing a circle of metastable states in the parameter space, which are characterized by the same depth of the scalar potential function.
(3) In order that there is a high probability for the transition F$\rightarrow$S, in the region where free flow is a metastable state, the congested state is either an unstable state, or it is a metastable state but geometrically difficult to be reached directly from the free flow state in the parameter space.  
In both cases, the chance to observe a direct F$\rightarrow$J transition becomes impossible or insignificant. In our model, the first possibility is adopted.
(4) The vector potential introduces oscillatory behavior in the F$\rightarrow$S transition process.

\begin{figure}[!htb]
\begin{tabular}{ccc}
\begin{minipage}{130pt}
\centerline{\includegraphics*[width=4.5cm]{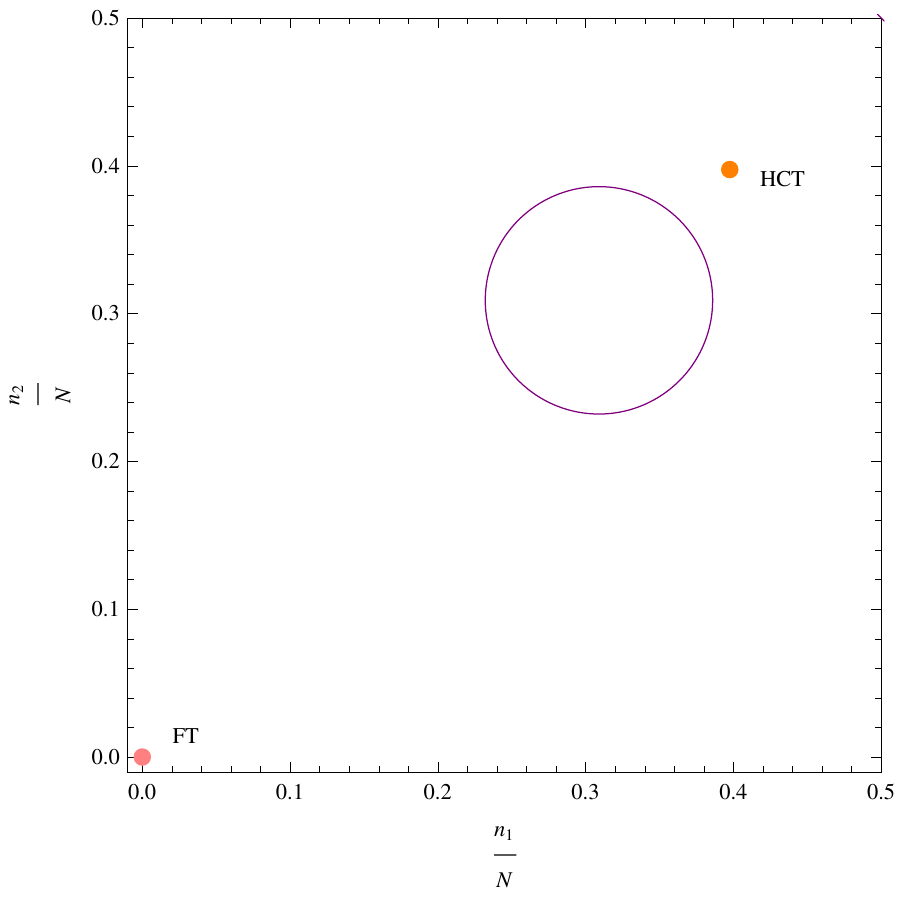} }
\end{minipage}
&
\begin{minipage}{130pt}
\centerline{\includegraphics*[width=4.5cm]{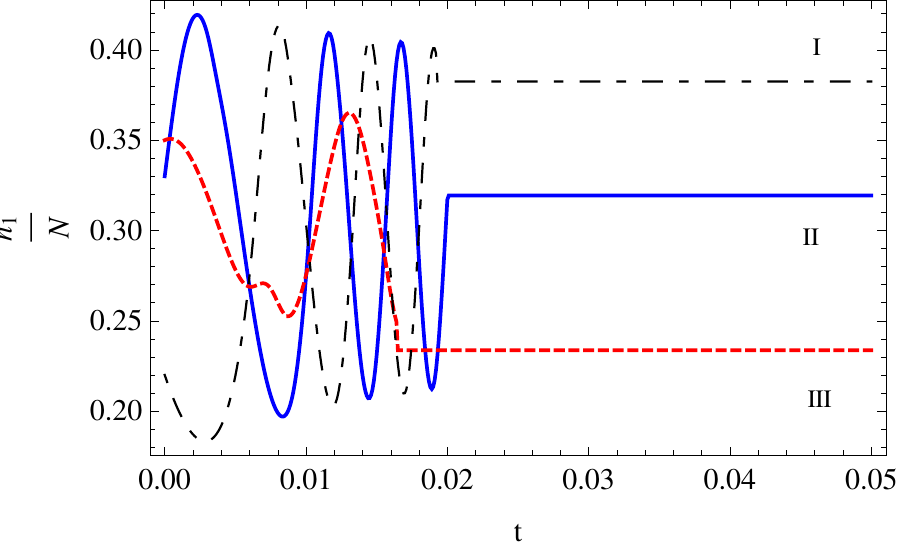} }
\end{minipage}
&
\begin{minipage}{130pt}
\centerline{\includegraphics*[width=4.5cm]{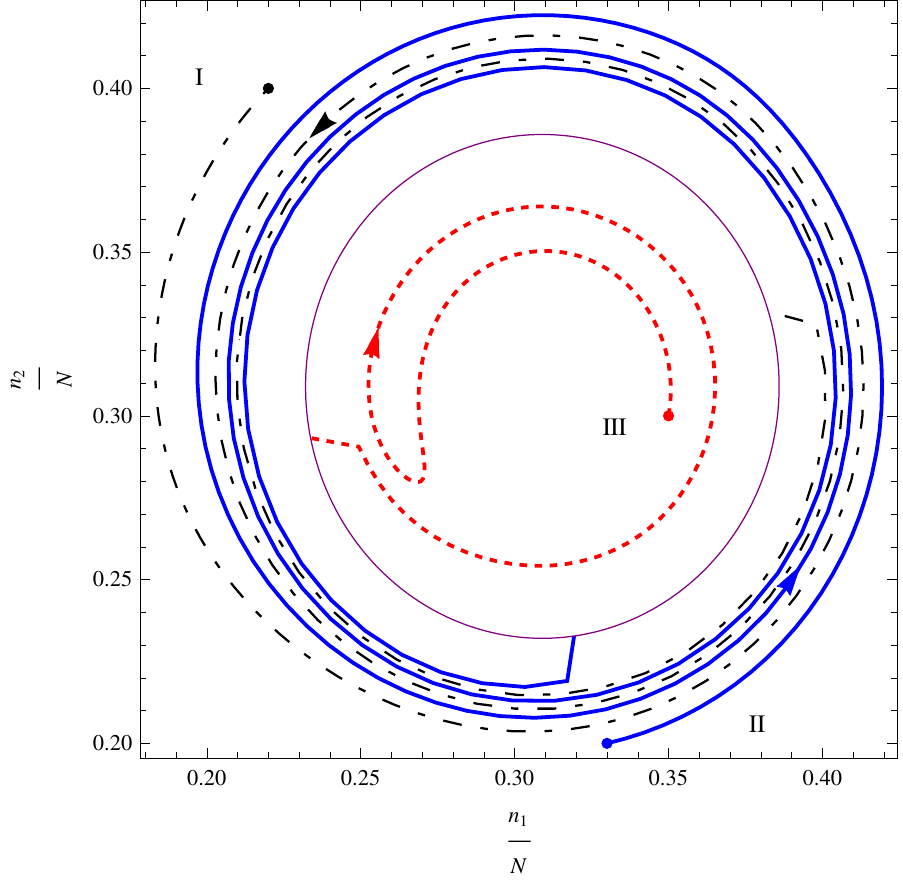} }
\end{minipage}
\\
\end{tabular}
\caption{(Color Online) Metastable region of the potential function and temporal evolution of the system for the second scenario of the two dimensional case.
Left: the minima of the potential function in the parameter space; the synchronized flow correspond to the circular region contains an infinite number of states, while other point-like stable states such as FT and HCT are also marked on the plot for comparison;
Middle: Temporal evolution of one of the parameters of the system. Depending on the initial conditions, the system oscillates and converges to any point in the metastable region where I II and III indicate three different initial conditions;
Right: The same as the middle plot, the evolution of the system is shown in the configuration space. 
}
\label{ev2d2}
\end{figure}

Again, the specific forms of the potential functions are relegated to the Appendix,
we only note that the form of the potential is chosen so that the inverse-$\lambda$ shape of the fundamental diagram (Fig.\ref{fdiagram}(c)) remains at the same position as in previous cases. 
For simplicity, the region of metastable states is taken to be a closed circle in the parameter space of the model, as illustrated in Fig.\ref{ev2d2}(a).
In Fig.\ref{ev2d2}(b) and (c), we show the temporal evolution of the system for different initial conditions in the synchronized flow phase. 
As shown in Fig.\ref{ev2d2}(b), depending on the initial condition, it is shown that the resulting state can pick any one from the family of metastable states shown in Fig.\ref{ev2d2}(a). If some small fluctuation is introduced, the system will be perturbed but evolves and eventually falls into another metastable state. 
In Fig.\ref{fdiagram}(c), we show the corresponding fundamental diagram. In the plot, the region featuring the scattering of the data is due to the region of metastable states. 
In terms of the fundamental diagram, it seems that both scenarios capture the feature of the scattered data. 
Fig.\ref{fdiagram}(c) also shows clearly that it is impossible for the system to evolve directly from a free flow state to the congested state. 
This is because for a given vehicle concentration, namely, a vertical line, will never intersect with both the curve representing the free flow state and that of the congested state.

\section{V. Concluding remarks}

In this work, a mesoscopic transport model is employed to study different traffic phases and their transitions.
Our approach is based on a mesoscopic gas-kinetic viewpoint of the traffic system in the same sense as proposed by Prigogine and Andrews \cite{traffic-flow-btz-01}.
By defining the potential function in the catastrophe model, the present work establishes an intuitive explanation for different states in traffic flow and their connection to the stability of the corresponding EoM. 
Our approach bears close connection to the catastrophe model and Landau's theory of phase transition. 
In the one-dimensional case, by adopting an appropriate form of the potential function, the model is able to reproduce stable steady traffic phases, including FT and HCT.
It reproduces the inverse-$\lambda$ feature in the fundamental diagram. 
When generalized to the higher-dimensional case, it is shown that the model provides a mechanism to describe the temporal oscillation in configuration space.
In our interpretation, the OST state is attributed to the formation of a limit circle; the synchronized flow state is attributed to a finite region of metastable states of the potential function. 
Therefore, elaborating such a dynamical theory not only provides a possible microscopic origin behind the potential function, but also makes it feasible to intuitively deal with phenomena such as stationary as well as non-stationary stable states.
The introduction of the potential function helps not only to quickly identify the locations of metastable states as well as oscillatory states, but also to quantitatively investigate the stability of such states.

As mentioned above, the topic of traffic flow states and their transitions have been explored by many authors. 
Most of the works employ either macroscopic hydrodynamics or microscopic car-following models.
Such approaches are usually based on detailed observations including traffic data and/or realistic highway and vehicle parameters.
Due to such sophistication, heavy numerical simulations are usually required where a thorough scanning of the model's parameter space as well as the initial conditions become increasingly difficult.
In addition, specific details in the construction of the model lead to delicate differences in the interpretation of numerical results, especially when the resulting solutions are not unambiguously identified in the observed data.
On the other hand, in the present study, we aim to develop an approach from a different angle.
Instead of focusing on specific details in modeling the traffic system, we concentrate on how the resulting traffic states are connected to the concept of instability and phase transition.
Inspired by the catastrophe model, all possible stable steady traffic states are associated with the local minima of the scalar potential, while vector potential gives rise to possible oscillatory traffic states.
The simplified mesoscopic gas-kinetic theory introduced in this work is to show the underlying physics more transparently. 
In particular, most of our discussions are from a phenomenological viewpoint. 
In this context, many findings of the current study are not dependent on any specific traffic scenario, neither on the particular external source in the EoM nor specific boundary conditions.
It is an impending task to further apply the present model to the description of the data.

In the above higher-dimensional models, the wide scattering of traffic data in the fundamental diagram is either interpreted in terms of the OST phase corresponding to a stable limit circle, or in terms of the synchronized flow phase corresponding to a region of metastable states.
For the first scenario, the OST phase is usually obtained by introducing pulse-type perturbations \cite{traffic-flow-hydrodynamics-15,traffic-flow-hydrodynamics-13,traffic-flow-hydrodynamics-12,traffic-flow-micro-17} into an initially homogeneous metastable state.
In our approach, coexisting solutions are obtained by choosing the initial conditions (as shown in the right column of Fig.\ref{ev2d1}) once the vector potential is properly given.
As we understand, both approaches effectively {\it kick} the system into a certain region of its configuration space, therefore bear the same physical content.
However, the cause of the limit circle is attributed to the curl of the potential function rather than to the inhomogeneity.
In our model, once the shape of potential functions are determined, it is relatively easy to find out what specific initial condition may lead to limit circle solution.
It is understood in \cite{traffic-flow-hydrodynamics-15} that oscillatory traffic state far away from the ramps are transient process of decaying oscillations.
In terms of our interpretation, this happens when the depth of the minimum of the corresponding scalar potential is so big that the
limit circle nearby becomes unstable. Therefore such phenomenon can be naturally described within the same framework. 
For the second scenario, the synchronized flow phase is obtained by introducing a region of metastable states, in the same sense as implemented in Ref.\cite{traffic-flow-cellular-06}.
According to the three phase traffic theory \cite{traffic-flow-three-phase-01,traffic-flow-three-phase-02}, the highway capacity is not a well-defined physical quantity. 
To have a region of metastable states implies that different values of highway capacity for different highway capacity, therefore it is consistent with the above assertion.
In addition, as shown in Fig.\ref{fdiagram}(c), our approach adopts an adaquate set of parameters that a direct transiton from free flow to congested flow becomes impossible. 
Comparing to Fig.\ref{fdiagram}(b), this can be achieved by decreasing the range of metastable congested flow. This can be done without much difficulty since the depth of scalar potential is easily manipulated in our model. In fact, by modifying the depth of corresponding metastable states, one can further adjust the probability of traffic break down as well as of the transition rates between different traffic states in accordance with experimental observations.

As discussed before, there are two possible sources of instability. 
This work only involves one of them, namely, the stability of the deterministic EoM and its relation to the traffic flow.
However, as traffic flow is stochastic in its nature, and our model itself involves a mathematical description of uncertainty,
it is intuitive to attribute the observed randomness in the data to the stochastic noises.
Such randomness, as we understand, can naturally provide a perturbation that takes place in reality to trigger the transition between different traffic phases.
Besides, with the presence of noises, phenomenon such as {\it hysteresis} effect \cite{traffic-flow-data-06} is expected to be reproduced.
In fact, the problem of stability of SDE has always been a topic of increasing attention in its own right \cite{stochastic-difeq-stability-01,stochastic-difeq-stability-02,stochastic-difeq-stability-03,stochastic-difeq-stability-04}.
It is meaningful to further construct a more realistic version of the model with its parameters determined by the observed characteristics of the traffic system.
We note that a preliminary attempt was made recently \cite{traffic-flow-btz-lob-04}.
The mesoscopic model for traffic flow has also been used to derive the corresponding macroscopic hydrodynamical equations \citep{traffic-flow-btz-03,traffic-flow-hydrodynamics-06,traffic-flow-hydrodynamics-08}.
The same procedure can be readily borrowed to derive the corresponding hydrodynamical equations and carry out studies of the traffic flow system from a macroscopic viewpoint as the latter is closely associated with experimental observables.
It is interesting to further investigate and compare the stability of the traffic flow states in both approaches. 
Do the spatial-temporal properties of the obtained hydrodynamical model reflect realistic characteristics of the traffic system? 
Moreover, are they related to those of the corresponding Boltzmann equation?
These are worthy topics for further exploration.

\section{VI. Acknowledgements}

We are thankful for valuable discussions with Jun Yan, De-Cheng Zou, Rong-Gen Cai, Hepeng Zhang, Olivera Miskovic, Rodrigo Olea, Ted William Grant and Adriano Siqueira.
This paper is greatly benefitted from the hospitality of Institute of Theoretical Physica of Chinese Academy of Science and Congqing University of Posts and Communication where most of the manuscript was written and intensively discussed.
We acknowledge the financial support from Funda\c{c}\~ao de Amparo \`a Pesquisa do Estado de S\~ao Paulo 
(FAPESP), Funda\c{c}\~ao de Amparo \`a Pesquisa do Estado de Minas Gerais (FAPEMIG),
Conselho Nacional de Desenvolvimento Cient\'{\i}fico e Tecnol\'ogico (CNPq),
and Coordena\c{c}\~ao de Aperfei\c{c}oamento de Pessoal de N\'ivel Superior (CAPES).

\section{VII. Appendix: The potential functions employed in the work}

For the first scenario of the two-dimensional case, the scalar potentials shown in Fig.\ref{pont2d2} read
\begin{eqnarray}
U^{(2)}\left(\frac{n_1}{N},\frac{n_2}{N}\right)&=&U^{(2)}_1\left(\frac{n_1}{N},\frac{n_2}{N}\right)+U^{(2)}_2\left(\frac{n_1}{N},\frac{n_2}{N}\right)+U^{(2)}_3\left(\frac{n_1}{N},\frac{n_2}{N}\right)\nonumber\\
U^{(2)}_i\left(\frac{n_1}{N},\frac{n_2}{N}\right)&=&d_i^{(2)}G_2\left(\frac{n_1}{N},\frac{n_2}{N},x_i,y_i,\sigma_i\right)~~~~(i=1,2,3)\nonumber\\
\label{pot-s-2d}
\end{eqnarray}
While the vector potential in Eq.(\ref{poteq2}) is defined as
\begin{eqnarray}
\vec{A}^{(2)}&=& A^{(2)}\hat{e}_z\\
A^{(2)}\left(\frac{n_1}{N},\frac{n_2}{N}\right)&=&d_r^{(2)}\sqrt{(x-x_r)^2+(y-y_r)^2}G_2\left(\frac{n_1}{N},\frac{n_2}{N},x_r,y_r,\sigma_r\right)
\label{pot-v-2d}
\end{eqnarray}
where the scalar and vector potentials are defined in terms of the ratio between occupation density of the two low velocity states $n_1$ $n_2$ and the total vehicle density $N$, 
$\hat{e}_z$ points in the direction perpendicular to the parameter space.
The parameters, the Gaussian function $G_2$ and Kernel function $K_2$ are defined as follows
\begin{eqnarray}
& &d_1^{(2)}=-0.01K_2(N,N,0,0,0.4),\nonumber\\
& &d_2^{(2)}=-0.0007K_2(N,N,0.7,0.7,0.14),\nonumber\\
& &d_3^{(2)}=-0.05K_2(N,N,1,1,0.6),\nonumber\\
& &d_r^{(2)}=700 d_3^{(2)},\nonumber\\
& &x_1=y_1=0,~~~\sigma_1=0.05,\nonumber\\
& &x_2=y_2=\frac{1}{4},~~~\sigma_3=0.03,\nonumber\\
& &x_3=y_3=\frac{1+N}{4},~~~\sigma_2=0.04,\nonumber\\
& &x_r=y_r=\frac{3}{20}+\frac{1+N}{10},~~~\sigma_r=0.5,\nonumber\\
& &\sigma = 0.1
\label{param-s-1d}
\end{eqnarray}
with
\begin{eqnarray}
G_2(x,y,x_0,y_0,\sigma)=\frac{1}{\sqrt{2\pi}}\exp\left(-\frac{(x-x_0)^2+(y-y_0)^2}{2\sigma^2}\right) \nonumber \\
\label{def-f2}
\end{eqnarray}
and
\begin{eqnarray}
K_2\left(x,y,x_0,y_0,h\right)=K_a\left(h^{-1}\sqrt{(x-x_0)^2+(y-y_0)^2},h\right)
\end{eqnarray}
\begin{equation}
K_a(q,h)=\left\{
  \begin{array}{lll}
    \frac{1}{\pi h^3}\left(1-\frac{3}{2}q^2+\frac{3}{4}q^3\right) & ~~~~~ & 0\le q\le1\\
    \frac{1}{4\pi h^3}\left(2-q\right)^3 & ~~~~~ & 1\le q\le2\\
    0 & ~~~~~ & 2\le q\\
  \end{array}
\right.
\end{equation}

As discussed in the text, the scalar potential which features three minima, corresponding FT, HCT and OST states. 
In our calculations, for simplicity, the maximal value of $N$ is scaled to be $1.0$.
The form of the minima is produced by a Gaussian function.
On the one hand, the minimum $x_1$ is related to the FT at low vehicle density, its position is fixed in the parameter space so that flow-concentration curve has a constant inclination for the FT phase as shown in Fig.\ref{fdiagram}(b). 
As discussed in section III, when the overall vehicle density increases, the minimum of FT starts to become metastable and eventually disappears.
This is controlled by the Kernel function $K_2$ defined above.
Similarly, the position of minimum $x_3$ corresponding to HCT moves towards $0.5$ (as $(n_1+n_2)/N$ moves towards $1.0$) as $N$ increases, so that more and more vehicles transit to the low velocity state $n_1$ and $n_2$ which consequently causes the overall traffic flow to decrease.
Since the two minima satisfy $x_3 > x_1$ in the region where FT and HCT coexist, the congested flow is always smaller than the free flow, and the difference corresponds to the jump as shown in Fig.\ref{fdiagram}.
On the other hand, the additional local minimum $x_2$ is introduced here to form the OST phase, together with the localized rotational field $\vec{B}^{(2)} = \nabla \times \vec{A}^{(2)}$ defined in terms of the vector potential $\vec{A}^{(2)}$.
Since the Kernel function $K_2$ is centered at $N=0.7$, the strength of the vector field $\vec{B}$ is also a function of the total vehicle density $N$, which is suppressed at lower and higher total vehicle densities.  
It is straightforward to verify that the vector field $\vec{B}$ forms closed circles around $(x_r,y_r)$ with an effective range of $\sigma_r$. 
It is noted that even though the equation of state of our two dimensional model is defined to be symmetric between $n_1$ and $n_2$,
it solution, as shown in Fig.\ref{ev2d1}, depending on the choice of initial conditions, does not necessarily possess such symmetry.
To numerically evaluate the resulting traffic flow, one assumes for simplicity that the two low velocity states are almost degenerate, 
namely, $v_1 \simeq v_2 \simeq 0$, while the higher velocity $v_3 = 1$.

The one-dimensional case can be obtained by removing the vector potential as well as the minimum corresponding to the OST state, and also considering only one unique low velocity state (perfect degeneracy) $n_1$. 
Due to the following intentional choices of the parameters, the one-dimensional model can be simply seen as a special case of degeneracy of its more sophisticated two-dimensional correspondence, and therefore corresponding flow-concentration relation for FT and HCT are exactly the same in the two plots in Fig.\ref{fdiagram}.
To be specific, in the one dimensional case, the scalar potential is defined as
\begin{eqnarray}
U^{(1)}\left(\frac{n_1}{N}\right) &=& U^{(2)}\left(\frac{n_1}{N},\frac{n_2}{N}\right) \nonumber \\
\label{pot-s-1d}
\end{eqnarray}
where
\begin{eqnarray}
d_2^{(2)} &=& 0,\nonumber\\
d_r^{(2)} &=& 0,
\end{eqnarray}
and all other parameters remain unchanged. 
Therefore the two minima $x_1$ and $x_3$ are interpreted as FT and HCT states, 
in exactly the same way as in the above two-dimensional case. 

For the second scenario of the two-dimensional case, the scalar potentials are essentially the same as those of the first scenario, an additional region of metastable states is introduced in the following:
\begin{eqnarray}
U^{(2)}\left(\frac{n_1}{N},\frac{n_2}{N}\right)&=&\left[U^{(2)}_1\left(\frac{n_1}{N},\frac{n_2}{N}\right)+U^{(2)}_2\left(\frac{n_1}{N},\frac{n_2}{N}\right)+U^{(2)}_3\left(\frac{n_1}{N},\frac{n_2}{N}\right)\right]F_{0.005}\nonumber\\
&&-(F_{0.005}-1)d_r^{(2)}\nonumber\\
U^{(2)}_i\left(\frac{n_1}{N},\frac{n_2}{N}\right)&=&d_i^{(2)}G_2\left(\frac{n_1}{N},\frac{n_2}{N},x_i,y_i,\sigma_i\right)~~~~(i=1,2,3)\nonumber\\
\label{pot-s-2d}
\end{eqnarray}
While the vector potential in Eq.(\ref{poteq2}) is defined as
\begin{eqnarray}
\vec{A}^{(2)}&=& A^{(2)}\hat{e}_z\\
A^{(2)}\left(\frac{n_1}{N},\frac{n_2}{N}\right)&=&F_{0.02}d_r^{(2)}\sqrt{(x-x_r)^2+(y-y_r)^2}G_2\left(\frac{n_1}{N},\frac{n_2}{N},x_r,y_r,\sigma_r\right)
\label{pot-v-2d}
\end{eqnarray}
where the parameters are as follows
\begin{eqnarray}
& &F_{\alpha}=1-\frac{G_2\left(\sqrt{(x-x_r)^2+(y-y_r)^2},\sqrt{(x-x_r)^2+(y-y_r)^2},x_\sigma,x_\sigma,\alpha\right)}{G_2\left(x_\sigma,x_\sigma,x_\sigma,x_\sigma,\alpha\right)},\nonumber\\
& &d_1^{(2)}=-0.01K_2(N,N,0,0,0.4),\nonumber\\
& &d_2^{(2)}=-0.003K_2(N,N,0.6,0.6,0.25),\nonumber\\
& &d_3^{(2)}=-0.05K_2(N,N,1,1,0.6),\nonumber\\
& &d_r^{(2)}=700 d_3^{(2)},\nonumber\\
& &x_1=y_1=0,~~~~\sigma_1=0.05,\nonumber\\
& &x_2=y_2=\frac{1}{4},~~~\sigma_3=0.03,\nonumber\\
& &x_3=y_3=\frac{1+N}{4},~~~\sigma_2=0.04,\nonumber\\
& &x_r=y_r=\frac{3}{20}+\frac{1+N}{10},~~~\sigma_r=0.5,\nonumber\\
& &\sigma=0.1,~~~x_\sigma=1/13 \label{param-s-1d}
\end{eqnarray}

According to our mesoscopic flow, the traffic flow can be therefore obtained in terms of the expected value 
\begin{eqnarray}
q = \sum_i n_i v_i
\end{eqnarray}
Thus the flow-concentration relation can be evaluated once the temporal evolution of $n_i$ is obtained numerically. 
It is noted that the soluction of the one dimensional model can simply be obtained by
substituting $(n_1+n_2) \rightarrow n_1$ and $n_3 \rightarrow n_2$ in the two dimensional case.

\bibliographystyle{h-physrev}

\bibliography{references_qian}{}

\end{document}